%% file: 11358.tex
\documentclass[structabstract]{aa}  
\usepackage{graphicx}
\usepackage{txfonts}
\usepackage{natbib}
\newcommand{\cmmthree}{\mbox{cm$^{-3}$}}

\newcommand{\kms}{\mbox{km\,s$^{-1}$}}

\newcommand{\msun}{\mbox{M$_{\odot}$}}

\newcommand{\microns}{\mbox{$\mu$m}}
\newcommand{\micron}{\mbox{$\mu$m}}

\newcommand{\chthreeoh}{\mbox{CH$_{3}$OH}}
\newcommand{\hcop}{\mbox{HCO$^+$}}
\newcommand{\hthirteencop}{\mbox{H$^{13}$CO$^+$}}

\newcommand{\htwoo}{\mbox{H$_2$O}}
\newcommand{\ntwohp}{\mbox{N$_2$H$^+$}}

\newcommand{\thirteenco}{\mbox{$^{13}$CO}}
\newcommand{\ceighteeno}{\mbox{C$^{18}$O}}
\newcommand{\nhthree}{\mbox{NH$_3$}}

\newcommand{\rarr}{\rightarrow}

\newcommand{\hii}{H{\scriptsize II}}
\newcommand{\uchii}{UCH{\scriptsize II}}
\def\degrees{\hbox{$^\circ$}}

\begin{document}
   \title{Multi-generation massive star-formation in NGC\,3576}


   \author{C. R. Purcell \inst{1,2}
     \and
     V. Minier \inst{3,4}
     \and
     S. N. Longmore \inst{2,5,6} 
     \and 
     Ph. Andr\'e \inst{3,4}
     \and
     A. J. Walsh \inst{2,7}
     \and
     P. Jones \inst{2,8}
     \and
     F. Herpin \inst{9,10}
     \and
     T. Hill \inst{2,11,12}
     \and
     M.~R.~Cunningham \inst{2}
     \and
     M. G. Burton \inst{2}}
   
   \institute{Jodrell Bank Centre for Astrophysics, Alan Turing
     Building, School of Physics and Astronomy, The University of Manchester, 
     Oxford Road, Manchester M13 9PL, UK.\\
     \email{cormac.purcell@manchester.ac.uk}
     \and
     School of Physics, University of New South Wales, Sydney, NSW 2052,
     Australia\\
     \and
     CEA, DSM, IRFU, Service dœôòùAstrophysique, 91191 Gif-sur-Yvette, France\\
     \and
     Laboratoire AIM, CEA/DSM - CNRS - Universit´e Paris Diderot, IRFU/Service dœôòùAstrophysique, CEA-Saclay, 91191
     Gif-sur-Yvette, France\\
     \and
     Harvard-Smithsonian Centre For Astrophysics, 60 Garden Street,
     Cambridge, MA, 02138, USA\\
     \and
     CSIRO Australia Telescope National Facillity, PO Box 76, Epping,
     NSW 1710, Australia\\
     \and
     Centre for Astronomy, James Cook University, Townsville, QLD
     4811, Australia\\ 
     \and
     Departamento de Astronomía, Universidad de Chile, Casilla 36-D, Santiago, Chile\\
     \and
     Universit\'e de Bordeaux, Laboratoire d'Astrophysique de Bordeaux, F-33000 Bordeaux, France\\
     \and
     CNRS/INSU, UMR 5804, BP 89, 33271 Floirac Cedex, France\\
     \and
     School of Physics, University of Exeter, Stocker Road, EX4 4QL,
     Exeter, UK\\
     \and
     Leiden Observatory, Leiden University, PO BOX 9513, 2300 RA Leiden, the Netherlands\\}

   \date{Received December 16, 2008; accepted July 3, 2009}

 
  \abstract
   {Recent 1.2-mm continuum observations have shown the giant
  H{\scriptsize II} region NGC\,3576 to be embedded in the centre of
  an extended filamentary dust-cloud. The bulk of the filament away
  from the \hii~region contains a number of clumps seen only at
  (sub-)millimetre wavelengths. Infrared and radio observations of the
  central star cluster have uncovered evidence of sequential
  star-formation leading us to believe that the adjacent clumps may
  host massive protostellar objects at a very early stage of
  evolution.}
   {We have investigated the physical and chemical conditions in the
     dusty clumps with the goal of characterising their star-forming content.}
   {We have used the Australia Telescope Compact Array (ATCA) to image
     the cloud for the \nhthree\,(1,1), (2,2) and (4,4)
     transitions, 22\,GHz \htwoo~masers, and  23\,GHz
     continuum emission. The 70-m Tidbinbilla dish was used to
     estimate the total integrated intensity of \nhthree. We also
     utilised the 22-m Mopra antenna to map the region for the
     molecular lines \thirteenco\,(1\,--\,0), \ceighteeno\,(1\,--\,0),
     \hcop\,(1\,--\,0), \hthirteencop\,(1\,--\,0), CS\,(1\,--\,0) and
     \ntwohp\,(1\,--\,0). }
   {Emission from dense molecular gas follows the morphology of the
     1.2-mm dust emission, except towards the central ionised region. The
     \hii~region is observed to be expanding into the molecular cloud,
     sweeping up a clumpy shell of gas, while the central star cluster is
     dispersing the molecular gas to the east. Analysis of the
     \nhthree~data indicates that temperature and linewidth
     gradients exist in the western arm of the filament. Temperatures
     are highest adjacent to the central \hii~region,
     indicating that the embedded cluster of young stars there is
     heating the gas. Six new \htwoo~masers were 
     detected in the arms of the filament, all associated with
     \nhthree~emission peaks, confirming that star-formation has
     begun within these cores. Core masses range from 5 to 516\,\msun~and most
     appear to be gravitationally bound. Complementary results by
     Andr\'e et al. (2008) imply that seven cores will go on to
     form massive stars between 15 and 50\,\msun. The large scale
     velocity structure of the filament is smooth, but at least one
     clump shows the signature of inward gas motions via asymmetries
     in the \nhthree\,(1,1) line profiles. The same clump exhibits an
     enhanced abundance of \ntwohp, which coupled with an absence of
     CO indicates depletion onto the dust grain surface. }
   {The H{\scriptsize II} region at the heart of  NGC\,3576 is
     potentially triggering the formation of massive stars in the bulk of the
     associated cloud.}

   \keywords{
     ISM:molecules --- stars:formation --- HII regions --- radio
     lines:ISM --- ISM:abundances --- surveys ---
     stars:pre-main-sequence }

   \maketitle
%

\section{Introduction}\label{sec:intro}
\subsection{Triggered star-formation}
A necessary precursor to massive star formation is the
existence of dense clumps of self-gravitating
gas. \citet{Elmegreen1998} developed the first coherent picture of
sequential star formation, in which the formation and collapse of
these clumps within giant molecular clouds is triggered by an external
event. Three distinct triggering mechanism are considered: 
\begin{enumerate}
  \item Globule squeezing: compression of pre-existing clumps,
  e.g. due to a propagating shock-wave from a supernova. 
  \item Cloud-cloud collisions: two molecular clouds collide resulting in
  gravitational instabilities. 
  \item Collect and collapse: accumulation of gas into a shell or
  ridge, and subsequent fragmentation and collapse.
\end{enumerate}

Examples of the third mechanism are observed to occur on the edges of
\hii~regions. Created by the far-UV radiation from a young OB-cluster,
the hot ionised gas expands rapidly, sweeping up a shell of dense
molecular material before it. A photon dominated region (PDR) is
created at the interface between the \hii~region and the molecular 
cloud (see \citealt{Hollenbach1999} for a review of PDRs). If the PDR
is over-pressured compared to the bulk of the molecular gas, shocks
are driven into the dense neutral medium of the cloud, potentially
leading to its fragmentation into clumps (e.g. \citealt{Urquhart2006}
and references therein). Subsequent disturbances may then trigger
collapse and hence star formation. Material swept into a clump shields
the column of molecular material behind it, forming giant pillars, the
best known example of which are the `elephant trunks' in the Eagle
Nebula \citep{White1999,Allen1999}. Some observed examples of swept-up
shells around \hii~regions include Sh2-212 \citep{Deharveng2008},
  RCW120 \citep{Deharveng2009}, W5 \citep{Karr2003}, RCW\,49
\citep{Whitney2004}, and RCW\,79 \citep{Zavagno2006}.

It is important to note that energy injected into molecular clouds by 
newly formed stars may instead quench the star formation
process. Supersonic particle-winds from massive stars account for
$\sim$\,0.1\,-\,1.0 per cent of the stellar luminosity and act to
disperse molecular material and dust (e.g., \citealt{Genzel1991}). The two most
important feedback mechanisms, photo-ionisation and cluster winds,
compete with each other to disrupt the cloud. On the scale of giant
molecular clouds, feedback from newly formed stars is responsible for
regulating the star-formation rate and hence the evolution of Galactic
structure.

In this paper we present new multi-wavelength observations of the
giant \hii~region NGC\,3576, which is thought to be undergoing
sequential star formation (see Section \ref{sec:prior_obs}). We
examine the evidence for star formation in the nearby  molecular
environment. 


\subsection{Prior observations of NGC\,3576}\label{sec:prior_obs}
\begin{figure*}
  \centering
    \includegraphics[width=17cm, trim=0 -0 0 0]{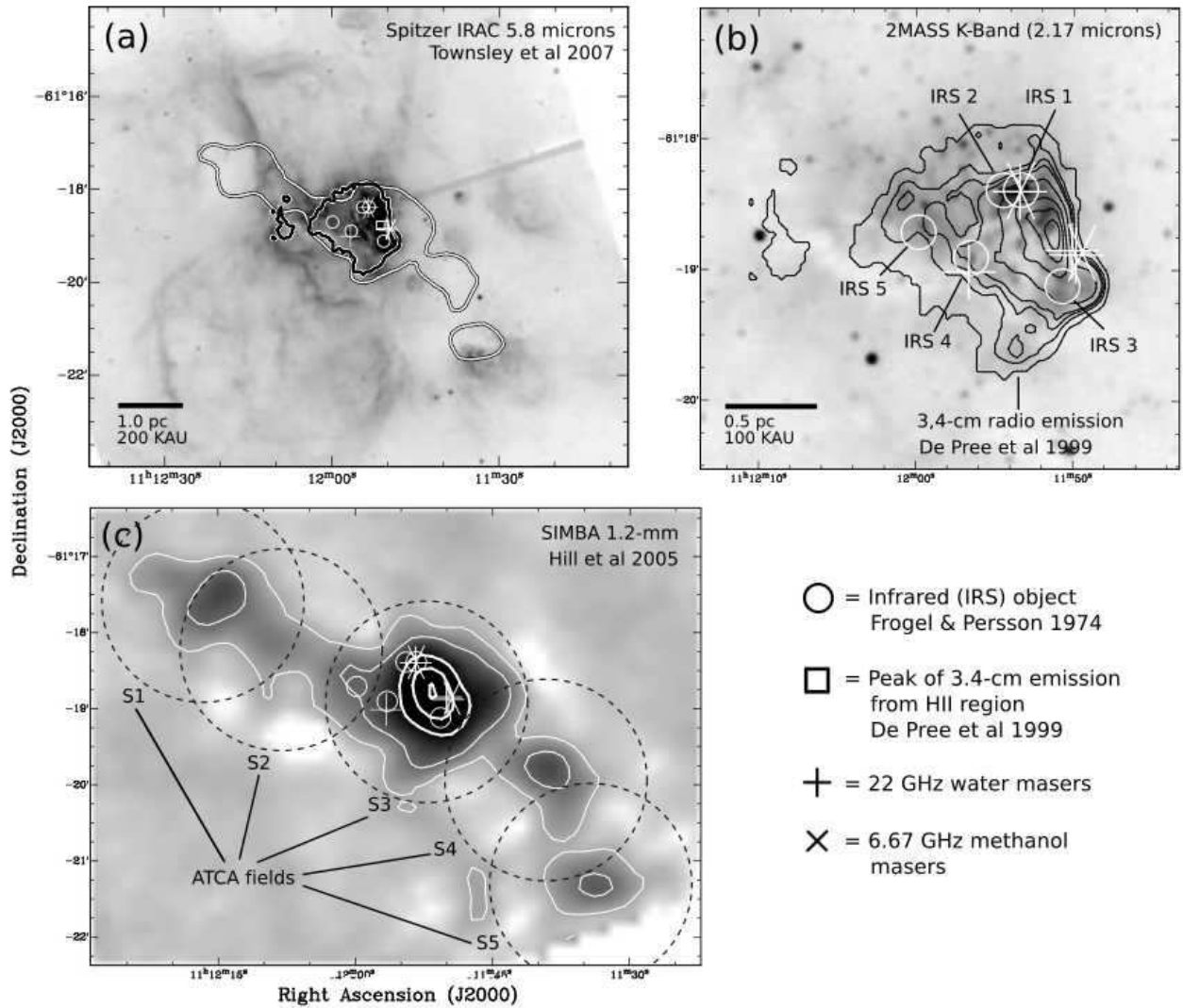}
    \caption[Overview of the NGC\,3576 star forming
    region.]{Overview of the NGC\,3576 star forming
    region. ({\bf a}) SPITZER 5.8\microns~image showing
    the extent of the 3.4-cm radio emission (black contour, $\sim$3~percent)
    and the 1.2-mm continuum emission (white contour, $\sim$5~percent).
    \chthreeoh~and \htwoo~maser positions are marked with `$\times$'
    and `$+$' symbols, respectively. Circles mark the positions
    assigned to the original infrared sources (IRS) discovered by
    \citet{Frogel1974}, while a `\,$\Box$\,' marks the peak of the
    radio-continuum emission. ({\bf b}) Contours of the 3.4-cm
    free-free  emission from ionised gas in the giant H{\scriptsize
      II} region at the heart of the complex \citep{DePree1999}
    plotted over the 2.2\microns~K-band image from 2MASS. ({\bf c})
    1.2-mm continuum image of the complex \citep{Hill2005} showing the
    dusty filament running across the H{\scriptsize II} region. Dashed
    circles indicate the five fields observed with the ATCA, centred
    on the 1.2-mm clumps.}
    \label{fig:summary}
\end{figure*}
The environment of the giant \hii~region NGC\,3576 (also known as
RCW\,57, G291.3$-$0.7 and IRAS\,11097$-$6102) has been extensively
studied at infrared and radio wavelengths. Figure~\ref{fig:summary}-a
presents the 5.8\microns~SPITZER IRAC image of the whole star-forming
complex with the \hii~region at the centre. The hourglass shaped
emission feature running north-south across the centre of the image
likely correspond to a bipolar cavity being 
evacuated by the central embedded cluster. The giant molecular cloud
which hosts NGC3576 manifests itself as large extincted areas  
to the north-east and south-west, and is at a distance of 2.4\,kpc
\citep{Persi1994}. The extent of the H{\scriptsize II} 
region is  illustrated by the thick black contour, corresponding to the
3-$\sigma$ (54\,mJy/beam) level in the 3.4-cm free-free emission
mapped by \citet{DePree1999}. Figure~\ref{fig:summary}-b shows the 
distribution of the 3.4-cm emission plotted over the
2.17\micron~infrared image from 2MASS. The ionised 
gas covers the brightest infrared emission, peaking sharply in
the west and extending $\sim$\,2\arcmin~to the
north-east and south. \citet*{Frogel1974} discovered five
near- and mid-infrared sources towards the ionised gas (plotted as circles
in Figure~\ref{fig:summary}), with the brightest source (IRS\,1)
located adjacent to the main peak of the radio emission. Further
high resolution photometric observations by \citet{Persi1994} have
revealed the presence of a very young, deeply embedded cluster with
130 members in the same region. Spectra of these sources exhibit deep
9.7\micron~silicate absorption features \citep{Moorwood1981},
leading to their interpretation as pre-main-sequence objects.

There is controversial evidence for sequential star formation in
NGC\,3576. \citet{Persi1994} showed that a steep near-IR colour
gradient exists in the embedded cluster, with the reddest, most deeply
embedded sources in the west. This implies that star-formation
began in the east, gradually moving to the location of the strong
radio peak in the west. Such an interpretation is supported by the
existence of an electron temperature gradient in the ionised gas
\citep{DePree1999} indicating that the youngest and hottest stars are
located near the sharp western edge of the \hii~region
\citep{Hjellming1966}. High helium abundance in the east may also 
indicate the presence of a population of older stars and their mass 
loss \citep{Hanson1993}. Bright 22\,GHz water masers, commonly found
in outflows, have been found adjacent to the main radio peak
\citep{Caswell2004}, while 
\citet{Norris1993} also detected two 6.67\,GHz methanol maser sites
near IRS\,1 and IRS\,3. Shocked molecular hydrogen emission was measured
peaking towards 
IRS\,1 by \citet{Oliva1986}, also pointing to the existence of
outflows. IRS\,1 was later resolved into three components by
\citet{Moneti1992} and \citet{Barbosa2003}, one of which is seen
through scattered light from a cavity or a dusty disk
\citep{Moneti1992}. None of the detected infrared sources contributes
significantly to the ionisation of the \hii~region,
\citep{Barbosa2003}, however, recent Chandra observations
\citep{Townsley2006} have revealed multiple deeply embedded hard X-ray
sources, which may provide the extra ionisation needed. 

The \hii~region was mapped at 1-mm wavelengths for continuum emission,
with a resolution of 1\arcmin, by \citet{Cheung1980}, who found that the
flux density ratio compared to 40-350\microns~was consistent with optically
thin thermal dust emission. More recently, \citet{Hill2005} used the
SIMBA\footnote{Sest IMaging Bolometer Array on the Sweedish ESO
  Submillimeter Telescope.} bolometer to map 1.2-mm continuum emission
at a resolution of $\sim$\,24\arcsec, as shown in
Figure~\ref{fig:summary}-c and via white contours on
Figure~\ref{fig:summary}-a. The SIMBA field covers an area of
4\arcmin$\times$\,6\arcmin, centred on the radio peak and reveals that
the \hii~region is embedded in a 
filamentary structure of cool dust, running north-east to
south-west. The H{\scriptsize II} region is prominent at the
centre, however, several bright knots are apparent along the length of 
the filament. These clumps, designated S1\,--\,S5 in this work, fall along
the dark lane traversing the nebulosity in the near-infrared image, and
are coincident with infrared-dark clouds (IRDCs) in the mid-infrared
(e.g., the 5.8\micron~IRAC image). IRDCs have been found throughout
the Galactic plane with masses in excess of 30\,\msun~and it has been
suggested they are the cold precursors to massive star clusters
(e.g., \citealt{Pillai2006}, \citealt{Rathborne2006}). 

NGC\,3576 constitutes an ideal laboratory in which to study the
process of massive star formation. In particular, we aim to
investigate if the \hii~region is interacting with the dusty
filament and if star-formation has been triggered in the dense knots
along its length. Here we present new observations of the
whole filament in several molecular tracers, with the goal of
constraining the star-forming properties of these new clumps (S1, S2,
S4 \& S5), and determining the effect of the H{\scriptsize II} region
on the complex.


\section{Observations and data reduction}
Data from three telescopes, the Australia Telescope Compact Array
(ATCA), the 22-m Mopra telescope and the 70-m Tidbinbilla telescope,
were combined to assemble a picture of NGC\,3576. The ATCA was used to
obtain high resolution \nhthree~maps as a probe of
the density and temperature structure of the clumps within the
filament. We simultaneously searched for \uchii~regions via their
23\,GHz free-free continuum emission. We used the
Tidbinbilla telescope to map the extended \nhthree~emission, providing
a measure of the total column of \nhthree~in the filament and an
estimation of the `missing flux' in the interferometer images. We
utilised the Mopra telescope to map the complex in the 3-mm lines
\thirteenco\,(1\,--\,0), \ceighteeno\,(1\,--\,0), \hcop\,(1\,--\,0),
\hthirteencop\,(1\,--\,0), \ntwohp\,(1\,--\,0) and CS\,(2\,--\,1),
with the goal of probing the physical and chemical conditions in each
of the dusty clumps. Line rest frequencies and electronic constants
for each transition are noted in Table~\ref{tab:transitions3} in the
Appendix.


\subsection{ATCA observations}
\input tables/11358tb1.tex
Observations were made with the ATCA in three blocks during the years
2003\,--\,2005. Table \ref{tab:atcaobs} summarises details of the
dates, array-configurations and frequencies used.
\nhthree\,(1,1) data was obtained using the EW367 and H75 array
configurations in August 2003 and July 2005, respectively. Two
orthogonal linear polarisations were observed at each frequency. The
raw data were processed in a correlator, which was configured to
deliver a bandwidth of $\sim$\,101\,\kms~split into
$\sim$\,0.2\,\kms~wide 
channels. \nhthree\,(2,2) was observed simultaneously with
\nhthree\,(1,1) on the H75 array in July 2005. Due to technical
constraints only 128 channels were available on the correlator,
yielding a spectral resolution of $\sim$\,0.8\,\kms~over the
101\,\kms~bandwidth. In August 2003 the region was also mapped for
23\,GHz continuum and 22\,GHz water masers using the  EW367
array. In July 2004, the 750D array was used to simultaneously 
map the \nhthree\,(4,4) transition and 23\,GHz continuum. The
correlator configuration was identical to that used to obtain the
\nhthree\,(1,1) data.

The primary beam of the ATCA at 23\,GHz has a full width half maximum (FWHM)
of 2.5\arcmin. In order to cover the extent of the 1.2mm continuum
emission we observed five overlapping fields, marked by dashed circles in
Figure~\ref{fig:summary}-d and centred approximately on the coordinates of the
dusty SIMBA clumps. Table~\ref{tab:ngc3576_coordinates} lists the
coordinates of the pointing centres. Each of the five fields was
observed for 10 minutes in turn, over the course of one hour and this
pattern was repeated for 10 hours, giving a total of 1.7 hours on each
position. In order
to correct for fluctuations in the phase and amplitude caused by
atmospheric and instrumental effects, a strong phase calibrator was
observed for two minutes before and after changing fields. The
instrumental contribution to the bandpass shape was measured by
integrating on a strong continuum source (e.g., 0420-014) and was
subsequently subtracted from all spectra. A primary flux calibrator
(1934-638 or Uranus) was observed once per observation period, to
allow the absolute calibration of the flux scale.  
\input tables/11358tb2.tex

The data were reduced using the {\scriptsize MIRIAD} package
\citep{Sault1995} following standard procedures. During the data
reduction the sources were assumed to be unpolarised and both
polarisations were averaged together. Continuum emission was
subtracted from the spectral line data by using the task {\it uvlin}
to fit a polynomial to the line free-channels. An image of the 
continuum emission was produced by combining the line-free channels in
the \nhthree~data with the dedicated wide-band 22 and 23\,GHz continuum
observations. Images were made using the {\it invert} task and natural
  weighting was used to minimise the noise in the image-plane. At this
  stage the $\sim$\,3\,km baselines to antenna CA06 
were discarded as the phases were found to be decorrelated, degrading
the image quality. All images were deconvolved using the standard {\it
  clean} algorithm and, if a sufficiently bright source was present,
several iterations of the {\it selfcal} task were applied. The
above procedure was performed on all fields, before using the task
{\it linmos} to mosaic the data into a single map. Finally the
\nhthree~data were converted to a brightness temperature scale in
Kelvin, using the Jy/K scaling factors noted in
Table~\ref{tab:atcaobs}.


\subsection{Mopra observations}
The Mopra observations were conducted in `on the fly' (OTF)
mapping mode, between the months June\,--\,September, during 2004 and
2005. An image was built up by combining overlapping scan rows, each
containing 30\,--\,40 spectra. Scan rows were offset by half of the
beam-FWHM and the scanning speed was slow enough to ensure Nyquist
sampling in the scan direction. 

NGC\,3576 was divided into an overlapping mosaic of three or more
5\arcmin\,$\times$\,5\arcmin~fields, positioned to cover the
1.2-mm continuum emission. The pointing centres of individual fields
varied, depending on where molecular emission was detected, and the
final maps were assembled by co-adding individual fields
together. Each field took $\sim$\,80 minutes to complete, plus a
further $\sim$\,10 minutes for pointing checks and calibration 
measurements. 

The signal from the receiver was processed in a digital
auto-correlator, configured to have a bandwidth of 64\,MHz divided
into 1024 channels, which provided a velocity resolution of
$\sim$\,0.2\,\kms~over a usable bandwidth of
$\sim$\,120\,\kms. The central frequency was chosen so that
channel 512 was centred on the systemic velocity of NGC\,3576 at
$-$24\,\kms. Observations were made in dual orthogonal linear
polarisation mode and the polarisations were averaged together during
the reduction procedure. The pointing accuracy was checked using a
nearby SiO maser before observing each field and was estimated to be better
than 8\arcsec. Calibration to the T$_{\rm A}^{\ast}$ scale was
achieved by measuring the emission from a single hot load placed in
front of the receiver every 20 minutes (see \citealt{Ladd2005} and
\citealt{KutnerUlich1981}). The maps were further calibrated onto the
main-beam brightness temperature scale (T$_{\rm MB}$) by dividing the
T$_{\rm A}^{\ast}$ pixel values by the main beam efficiency $\eta_{\rm
  mb}$ at the observing frequency (see
Table~4 of \citealt{Ladd2005}).

The data were reduced using the {\scriptsize LIVEDATA} and {\scriptsize
 GRIDZILLA} packages, available from the
ATNF\footnote{http://www.atnf.csiro.au/computing/software/}. {\scriptsize
  LIVEDATA}
performed bandpass calibration by subtracting the preceding OFF
spectrum from the SCAN spectra in each row. A low-order polynomial was
then fit to line-free channels and subtracted, resulting in a smooth
baseline at zero Kelvin. The spectra were assigned individual
position stamps and regridded to the LSR-K reference frame, before
being written to disk. The {\scriptsize GRIDZILLA} package was then used to
resample the maps to a regular pixel scale, weighted according to the
 system temperature (T$_{\rm sys}$). To grid the data we used a 
 pixel-size of  6$\times$6 arcseconds and a Gaussian smoothing kernel
 with a FWHM of 18\arcsec, truncated at an angular offset of
 36\arcsec. The final data cubes were smoothed to an angular
 resolution of $\sim$\,40\arcsec.


\subsection{Tidbinbilla observations}
The 70-m antenna located at the NASA Tidbinbilla Deep Space
Communications Complex near Canberra has a limited time devoted to
radio astronomy. When Mars is above the horizon the 70-m telescope is
almost entirely utilised supporting the various NASA missions to that
planet and as such has been little used by the radio astronomy
community. In 2005 we performed the first frequency-switched spectral
line mapping observations using the 23\,GHz K-band receiver. Fast frequency
switching also provides better cancellation of atmospheric emission
fluctuations, however, the response of the receiver may vary
considerably with frequency, resulting in poor spectroscopic
baselines.

We targeted the same five fields in NGC\,3576 as were observed with
the ATCA. The OTF method had not yet been implemented on the 70-m
antenna so we observed each field as a square grid of 5$\times$5
positions, spaced by half the 45\arcsec~beam-FWHM. Each position was
observed for 10 minutes and the pointing accuracy was checked by
observing an unresolved planet (Jupiter) every hour. Pointing errors were
usually below 10\arcsec~or 1/4 beam. Individual maps took 5\,--\,6
hours to complete, including 15 minutes to measure the
atmospheric opacity using the `skydip' method. During the observations
the data were calibrated against an ambient load (a noise diode) and
the zenith system temperature at 23\,GHz was typically 40\,--\,50\,K
during the observations. We estimate the T$_{\rm A}^{\ast}$ flux scale
is uncertain by  $\leq$\,10 per cent. The peak aperture efficiency
was measured as 0.48\,$\pm$\,5\%  at 22.2\,GHz by
\citet{Greenhill2003} and we adopt that value here to correct the data
onto the main beam brightness temperature flux scale (T$_{\rm
  MB}$). To convert to units of Jy/beam the data were multiplied by
0.951\,(Jy/beam)/K.

The K-band receiver on the 70-m antenna measured only the left
circular polarisation. The bandpass was centred mid-way
between the \nhthree\,(1,1) and (2,2) transitions and the correlator
was configured to have a 64\,MHz wide bandpass divided into 2048
channels. Both the \nhthree(1,1) and (2,2) spectra, including
satellite lines, fell well inside the 850\,\kms~usable range and were
observed at a velocity resolution of $\sim$\,0.4\,\kms.

The data were reduced in an ad-hoc pipeline built
using the {\scriptsize SPC}, {\scriptsize GILDAS} and {\scriptsize
MIRIAD} packages. {\scriptsize SPC} was used initially to quotient and
resample the spectra to a common V$_{\rm LSR}$ rest-frame. The spectra
were inverted and frequency-shifted in the {\scriptsize CLASS} package.
As the spectral baselines were poor, a polynomial of order 5\,--\,9
(in extreme cases) was fit to the line-free channels before spectra
from individual integrations were averaged together. The data were
then assembled into a 3D data-cube using the {\it cube} command in the
{\scriptsize GREG} package and resampled onto a finer spatial pixel
grid using the {\it fill\_cube} command. Finally, the data were output
as {\it fits} cubes and  {\scriptsize MIRIAD} was used to restore
missing header information, such as the beam-size and data-unit.


\section{Results}
Molecular emission was detected from all of the clumps identified in
the SIMBA map. The following subsections describe the results from the
ATCA, Tidbinbilla and Mopra telescopes separately.


\subsection{ATCA results}\label{sec:results_atca}
\begin{figure*}
  \begin{center}
    \includegraphics[angle=0, width=17.0cm, trim=-0 -20 0 0]{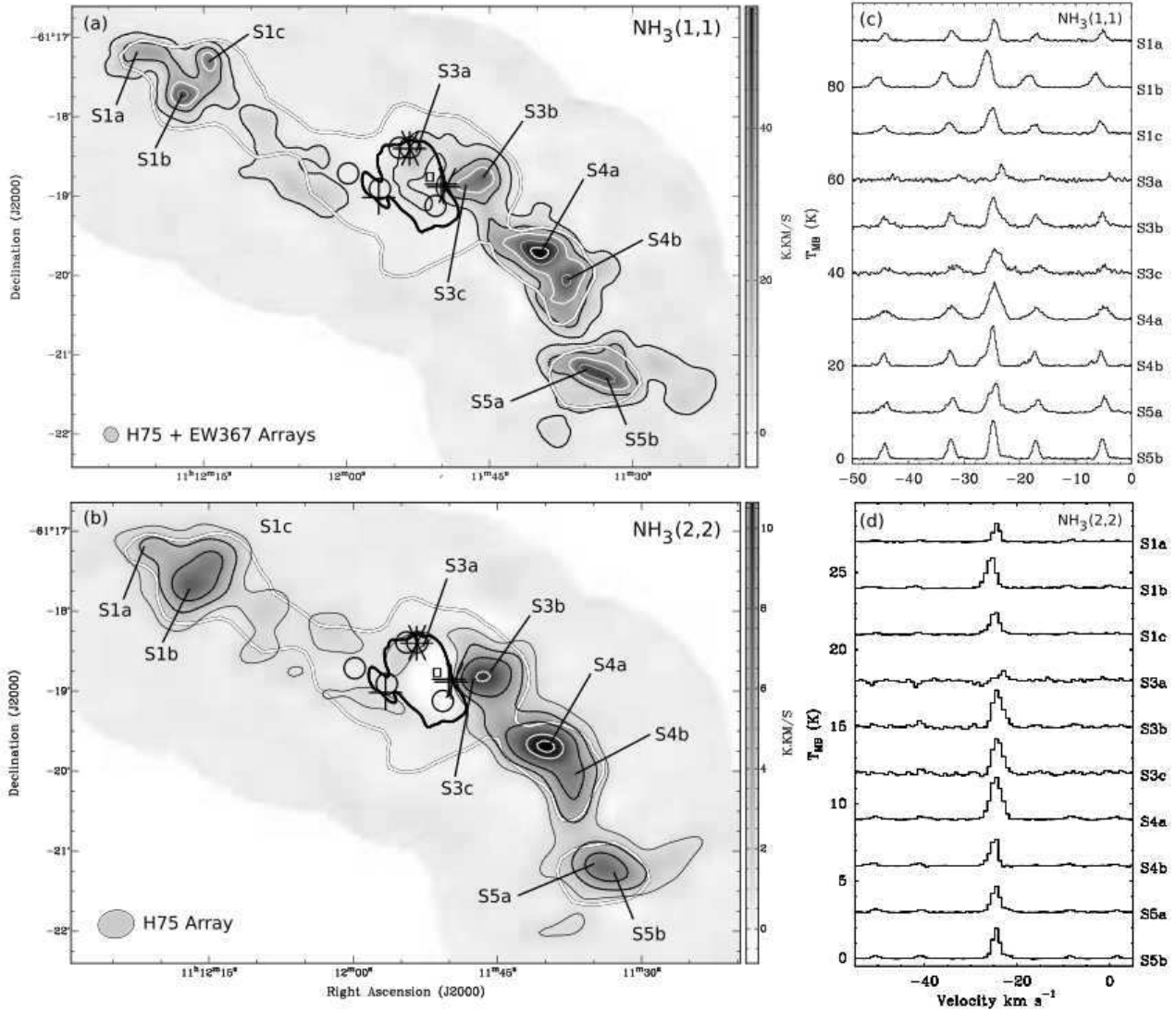}
    \caption[Integrated intensity maps and spectra of
      \nhthree.]{({\bf a}) Integrated intensity map of the \nhthree\,(1,1)
      emission (thin black and white contours, and greyscale) made
      using the combined ATCA H75 and EW367 data. Contours are at 1.6,
      7.9, 13.8, 20.0, 25.7, 31.5, 37.4,  43.6 and 50.1\,K. The extent
      of the 23\,GHz continuum emission from the central H{\scriptsize
      II} region is marked 
      by the thick black contour. The 1.2-mm dust emission is outlined
      by a single white contour ($\sim$\,3\,Jy/beam, from
      \citealt{Hill2005}) and matches the \nhthree~emission closely,
      except in the central clump. Crosses ($+$) mark the positions of
      22\,GHz water masers while `$\times$' symbols mark the 6.67\,GHz methanol
      maser sites. IRS sources are marked by circles. ({\bf b}) Integrated
      intensity map of the \nhthree\,(2,2) emission made using data
      from the ATCA H75 array. ({\bf c}) Sample \nhthree\,(1,1) spectra
      measured from the positions labelled
      S1a\,--\,S5b. Spectra have been offset from zero baseline for
      clarity. The brightness temperature ratios between
      the main and satellite lines indicate that the clumps have low
      to moderate optical depths. ({\bf d}) Sample \nhthree\,(2,2)
      spectra measured from the same positions. } 
    \label{fig:nh311_combined_integ}
  \end{center}
\end{figure*}
We detected \nhthree\,(1,1), and \nhthree\,(2,2) across the whole
filament. \nhthree\,(4,4) was not detected down to a sensitivity
limit of $\sim$\,8\,mJy/beam (1.0\,K) when the data were smoothed
to a velocity resolution of 1\,\kms. The J,K\,=\,(4,4) level is
excited by gas above 200\,K (see Figure~1 in \citealt{Ho1983}), hence
no substantial reservoirs of hot gas are present in
NGC\,3576. Figures~\ref{fig:nh311_combined_integ}-a and~-b present
the \nhthree\,(1,1) and (2,2) integrated intensity 
maps of the region. The \nhthree\,(1,1) map utilises data from both
the H75 and EW367 array configurations and has a resolution of
$\sim$\,11\arcsec. Only H75 data exist for the \nhthree\,(2,2)
transition and the resulting map has a resolution of
$\sim$\,23\arcsec. For reference, the 5 per cent contour
(0.3\,Jy/beam) of the SIMBA 1.2-mm continuum emission is plotted in
white. The thick black line is the 3 per cent contour (0.11\,Jy/beam)
of the 23-GHz continuum emission from the central
\hii~region. We see the \nhthree~emission has approximately the same
morphology in both transitions and closely follows the structure of
the 1.2-mm dust emission, except immediately to the east of the
central H{\scriptsize II} region (SIMBA clumps S2 \&
S3). The 1.2-mm emission in these clumps is over 50 percent
contaminated by free-free emission (see \citealt{Andre2008}). Sample
\nhthree\,(1,1) and (2,2) spectra  
from peak positions on the integrated intensity maps (labelled
S1a\,--\,S5b)  are presented in Figures~\ref{fig:nh311_combined_integ}-c
and -d. Both of the \nhthree~transitions exhibit the classic
`five-finger' profiles at all 
positions. Some asymmetries between the \nhthree\,(1,1)
satellite lines are evident towards all positions, however, for
the most part the spectra appear optically thin, exhibiting
  optical depths between $\tau=$\,0.1 and
  $\tau=$\,2.0. Linewidths vary between 1 and 2.5\,\kms, and peak
brightness temperatures range between  4.4 and 8\,K for the
\nhthree\,(1,1) line and between 1.2 and 2.6\,K 
for the \nhthree\,(2,2) lines. Individual line profiles in the western arm of
the filament show clear evidence of blending and multiple components,
especially towards positions S3b, S3c, S4b and S5a. 

\begin{figure}
  \centering
    \includegraphics[angle=0, width=8.0cm, trim=0 0 0 -20]{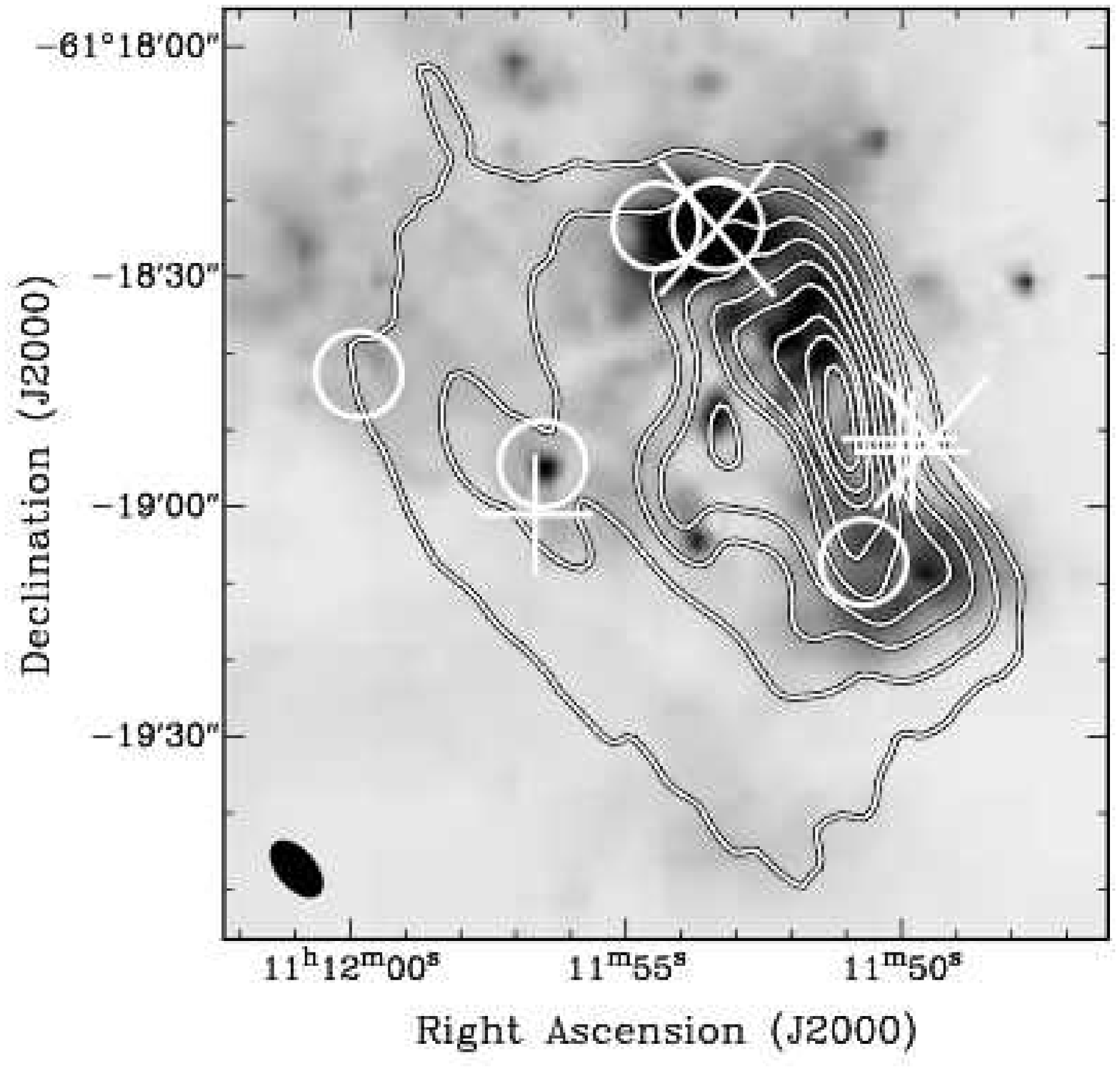}
    \caption[Map of 23-GHz continuum emission in NGC\,3576.]{23-GHz continuum emission from the
    H{\scriptsize II} region at the centre of the NGC\,3576 complex
    (contours) overlaid on the 3.6\micron~SPITZER IRAC image
    (greyscale). The continuum image was made by combining
    multi-frequency data from the ATCA H75, 750D and EW367 arrays,
    including the line-free channels from the
    \nhthree~observations. Contours are at values of 0.01, 0.11 0.45,
    0.85, 1.35, 1.65, 2.00, 2.48, 2.82, and 3.25 Jy/beam. `$+$' and
    `$\times$' symbols mark H$_2$O and \chthreeoh~masers,
    respectively, while circles mark the positions of the
    \citet{Frogel1974} IRS sources. The total integrated flux density
    is $\sim$\,42.6\,Jy.}
    \label{fig:22GHz_continuum}
\end{figure}
23\,GHz continuum was observed simultaneously with each of the
\nhthree~lines. To make the final map we combined data from the EW367,
H75 and 750D arrays, including the continuum visibilities from the
line-free channels in the narrow-band spectral line windows. No
23\,GHz continuum emission was detected
outside of the central H{\scriptsize II} region, down to a
sensitivity limit of
$\sim$\,0.5\,mJy/beam. Figure~\ref{fig:22GHz_continuum} presents the 
image of the H{\scriptsize II} region at
23\,GHz. `$\times$' and `$+$' symbols mark the positions of known 6.67\,GHz
methanol masers and 22\,GHz water masers, respectively
(\citealt{Norris1993}, \citealt{Caswell2004}). 
The contours are similar to the central part of the 3.4-cm map made by
\citet{DePree1999} (see Figure~\ref{fig:summary}-b), with a single strong peak at
11$^h$11$^m$51.08$^s$, $-$61$^d$18$^m$50.0$^s$ (J2000) and diffuse
emission extending eastward into SIMBA clump S2.  The 23\,GHz flux density at the peak is
3.6\,Jy/beam and we measure a total integrated intensity of approximately
42.6\,Jy. By comparison, the total flux at 3.4-cm is measured to be
71\,Jy. Our observations sample the {\it uv}-plane less well than
\citet{DePree1999} and we likely filter out a significant fraction of
the extended emission. Our values are consistent with free-free
emission from optically-thin ionised gas
(S$_{\nu}\,\propto\,\nu^{-0.1}$) if we are missing $\sim$\,34 percent
of the flux detected at 3.4-cm. 

\begin{figure*}
  \begin{center}
    \includegraphics[angle=0, width=14.0cm, trim=0 0 -0 0]{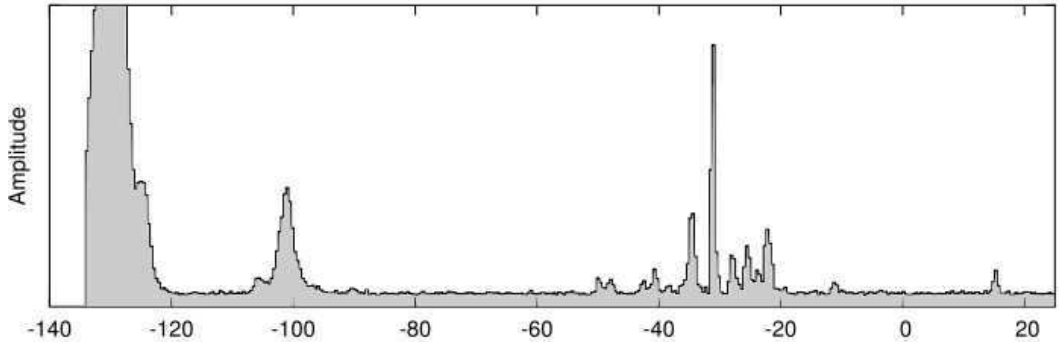}
    \caption[Cumulative H$_2$O maser spectrum.]{Cumulative
      H$_2$O maser spectrum from a single baseline showing lines from
    across the filament. The two broad maser lines at $-$130\,\kms~and
    $-$102\,\kms~derive from positions M9 and M5, respectively, towards
    the central \hii~region (see Table~\ref{tab:h2o_maser_coords}). The
    bulk of the maser lines concentrated near the systemic velocity at
    $-$24\,\kms~are scattered across the filament. The units of
    amplitude are arbitrary.}
    \label{fig:h2o_cumspec}
  \end{center}
\end{figure*}

\begin{figure*}
  \begin{center}
    \includegraphics[angle=0, width=15.cm, trim=0 0 -20 0]{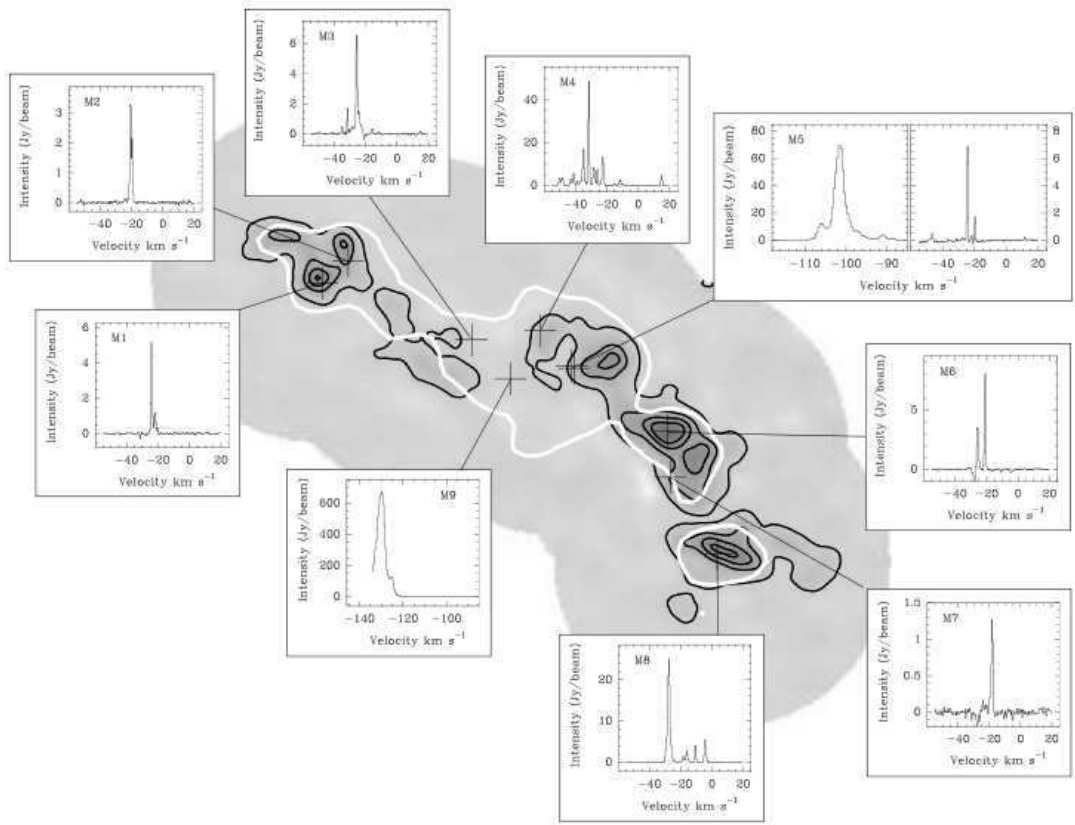}
    \caption[Positions and spectra for the 22\,GHz H$_2$O masers
    in NGC\,3576.]{Positions and spectra for the 22\,GHz H$_2$O masers
    detected in our ATCA EW367 array data. Six masers (M1, M2, M3, M6,
    M7 and M8) are new detections. Greyscale and black contours are
    the \nhthree\,(1,1) integrated intensity map. The white line is
    the $\sim$\,3\,Jy/beam contour from of the 1.2-mm continuum
    map. Individual spectra are sampled from the peak maser pixel in
    the {\it clean}ed images of five ATCA fields.} 
    \label{fig:h2o_masers}
  \end{center}
\end{figure*}

We also searched for 22\,GHz water maser emission across the
filament. New masers were detected at six positions, as well as at the
three previously known sites adjacent to 
IRS\,1, IRS\,3 and IRS\,4. Figure~\ref{fig:h2o_cumspec} presents a
cumulative spectrum from a single baseline showing the majority of
the maser lines detected. Two groupings of lines are observed, the
first centred at approximately $-$115\,\kms~(with two components at $-$130 and
-100km/s) and the second centred at $-$30\,\kms. The intense lines at
$-$130 and $-$100\,\kms~have peak amplitudes of $\sim$\,70 and
$\sim$\,900\,Jy/beam, respectively, and derive from maser sites towards
the central \hii~region. Maser sites scattered across the 1.2-mm
continuum filament contribute to the group of lines between $-$60 and
$-$10\,\kms. Figure~\ref{fig:h2o_masers} presents the maser spectra
and illustrates the positions of individual maser sites superimposed
on the \nhthree\,(1,1) zeroth-moment image. Precise coordinates, intensities and
velocities are noted in Table~\ref{tab:h2o_maser_coords}. 
\input tables/11358tb3.tex



\subsection{Tidbinbilla results}\label{sec:results_tid}
\begin{figure*}
  \begin{center}
    \includegraphics[angle=0, width=18.cm, trim=0 0 -20 0]{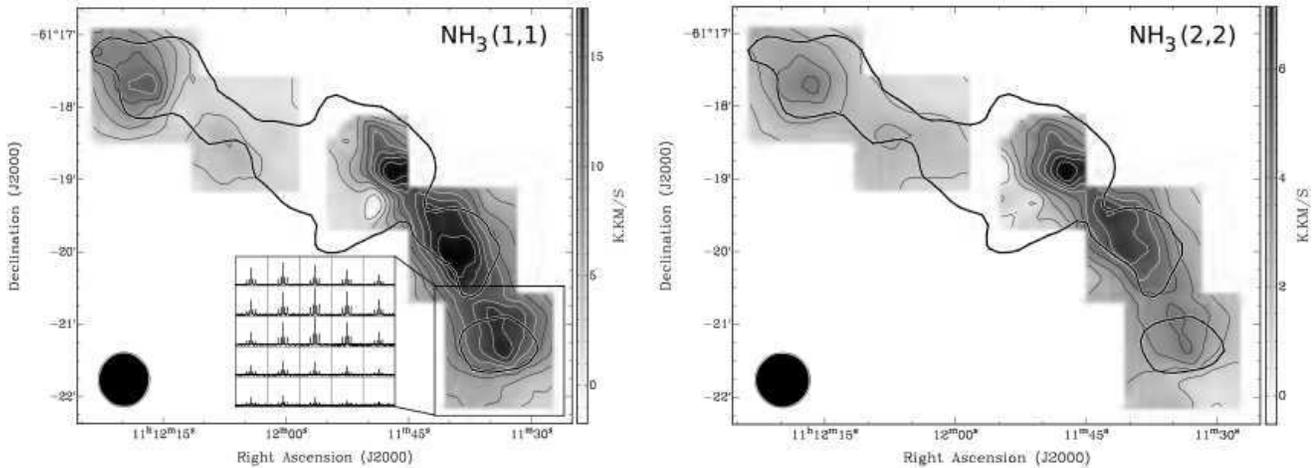}
    \caption[\nhthree~integrated intensity
    maps made using the Tidbinbilla
    telescope.]{\nhthree\,(1,1) and (2,2) integrated intensity 
    maps made using the 70-m Tidbinbilla telescope (greyscale and
    contours). Contours levels are at 10 percent intervals. The single
    thick black contour shows the extent of the 1.2mm continuum
    emission \citep{Hill2005}. The grid inset into the
    \nhthree\,(1,1) map illustrates the spectra which contributed to
    the map of the S5 field.} 
    \label{fig:nh3_tid_maps}
  \end{center}
\end{figure*}

\nhthree\,(1,1) and (2,2) integrated intensity maps made using
the Tidbinbilla 70-m telescope are presented in
Figure~\ref{fig:nh3_tid_maps}. Individual maps are centred on the same
position as the ATCA fields, however, technical difficulties meant
that we were unable to finish one column of field S3, covering the
\hii~region. It is immediately evident that the morphology of the
emission is similar to that seen in the the ATCA \nhthree\,(1,1)
data. 

We initially intended to use data from the 70-m antenna to `fill in'
the missing short-spacings in our ATCA data. For the merging to be
successful, the emission in the single-dish dataset must be imaged
out to its extremes. Unfortunately, there is still considerable flux
at the edge of all fields and, due to time constraints and the
experimental nature of the observations, we were unable to enlarge the maps.
Instead, the Tidbinbilla data is used to estimate the fraction of missing
flux due to extended emission in each of the ATCA fields. To do this we
smoothed the final ATCA maps to the same resolution and pixel scale as
the Tidbinbilla maps and compared the integrated intensities measured
under the same aperture. The percentage ATCA/Tidbinbilla integrated
intensity ratios are presented in
Table~\ref{tab:missing_flux}. Calibration error on the Tidbinbilla
data is approximately 30 percent and the measurements are consistent
with less than 10 percent missing flux in the ATCA fields.
\input tables/11358tb4.tex 
The estimates assume a beam-efficiency of 48 per cent for the 70-m
telescope \citep{Greenhill2003} and a (Jy/Beam)/K conversion factor of
0.951.


\subsection{Mopra results}\label{sec:results_mopra}
\begin{figure*}
  \centering
  \includegraphics[angle=00, width=18cm, trim=0 0 -0 0]{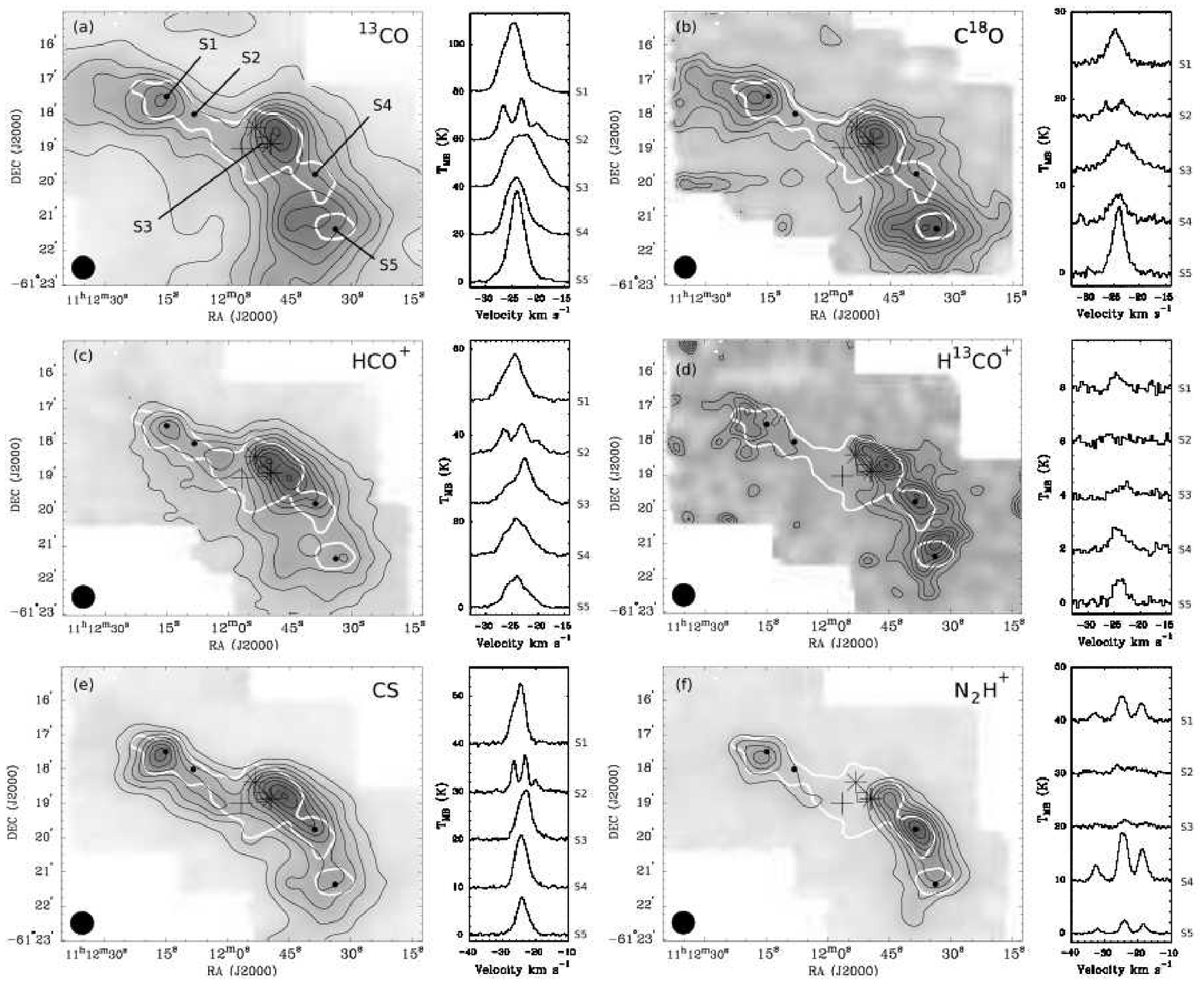}
    \caption{Integrated intensity (zeroth-moment) maps of 
    NGC\,3576 in the six molecular transitions observed with the Mopra
    telescope (left panels). Sample spectra, extracted from the data
    cube at the positions S1\,--\,S5, are presented to the right
    of each map. \thirteenco~is generally optically thick and is a good
    tracer of the extended molecular envelope.  \hcop~and
    \hthirteencop~trace similar gas densities to CO, while CS and
    \ntwohp~trace denser gas.}
    \label{fig:mopra_mom0}
\end{figure*}
Figure~\ref{fig:mopra_mom0} presents integrated intensity
maps and sample spectra for all six molecules detected by Mopra towards 
NGC\,3576. The data have been smoothed to a velocity
resolution of 0.4\,\kms~and have a spatial resolution of 40\arcsec. Sample
spectra presented alongside each map were extracted from the
data-cubes at the peak positions of the SIMBA clumps (labelled
S1\,--\,S5) identified in the 1.2-mm continuum map and marked by
filled circles in Figure~\ref{fig:mopra_mom0}.

As with \nhthree, the distribution of the other molecular lines  
closely follow the 1.2-mm emission except to the east of the central
H{\scriptsize II} region. \thirteenco\,(1\,--\,0) and
\hcop\,(1\,--\,0) trace more extended gas than the other lines, and
their spectra appear moderately 
optically thick compared to their isotopologues, \ceighteeno\,(1\,--\,0)
and \hthirteencop\,(1\,--\,0). \ntwohp~and CS both trace dense gas, however
\ntwohp~exhibits optically thin line profiles ($\tau <$\,0.1) at all
positions while CS appears optically thick in places. The relative
intensity of the emission between species varies significantly over
the extent of the cloud, likely due to differences in the chemistry of
the clumps. We discuss this possibility further in
Section~\ref{sec:chemistry}.

At position S2 the spectra of all species display
three line components, offset in velocity by $\sim$\,4\,\kms. Spectra
sampled at the other positions appear as a single line profile. These
lines are not well fit by single Gaussians except at position S5 and
are possibly composed of several blended features. Taking the
optically thin \ntwohp~as an example, the peak V$_{\rm lsr}$ of the
emission in the western arm (SIMBA clumps S3\,--\,S5) has a mean
velocity of $-$24.1$\pm$0.5\,\kms~, while the eastern arm (SIMBA clumps S1
and S2) has a mean velocity of $-$25.0$\pm$0.9\,\kms. No large velocity
gradients are apparent, but the two arms are offset in velocity by
$\sim$\,1\,\kms. The full-width half-maximum \ntwohp~linewidths ranges
between approximately 1 and 3\,\kms~across the filament, with the
highest values occurring adjacent to the \hii~region. 


\section{Derived Properties}

In this section we derive physical properties, such as temperature and
density, from the molecular line data. The detailed methodology used
in the calculations is presented in the appendix. We discuss the
results in Section 5 where we analyse the morphology, kinematics,
relative chemical abundances and star-forming content in detail. 

\subsection{Physical properties from \nhthree}
The rotation-inversion transitions of \nhthree~have been used
extensively in the literature to derive the kinetic temperature and
 column densities of molecular clouds (e.g.,
\citealt{Ungerechts1986,Cesaroni1992,Bourke1995}). Due to its quantum
mechanical properties, the \nhthree~inversion spectrum is split into
multiple hyperfine components from whose brightness ratios the optical
depth may be derived  directly, hence removing a major assumption from
the calculation of rotational temperatures. 


\subsubsection{Kinetic Temperature}
The modified rotational diagram method used to calculate rotational
temperature (T$_{\rm rot}$) from \nhthree~has been described in
detail by \citet{Ho1977}, \citet{Ungerechts1980} and
\citet{Townes1983}. We provide a brief summary of the procedure in the
appendix, specifically for the \nhthree\,(1,1) and (2,2)
transitions. The molecular constants used in the calculations are
collected in Table~\ref{tab:transitions3}.

\begin{figure*}
  \begin{minipage}{\textwidth}
    \centering
    \includegraphics[width=17.0cm, angle=0, trim=0 -20 0 0]{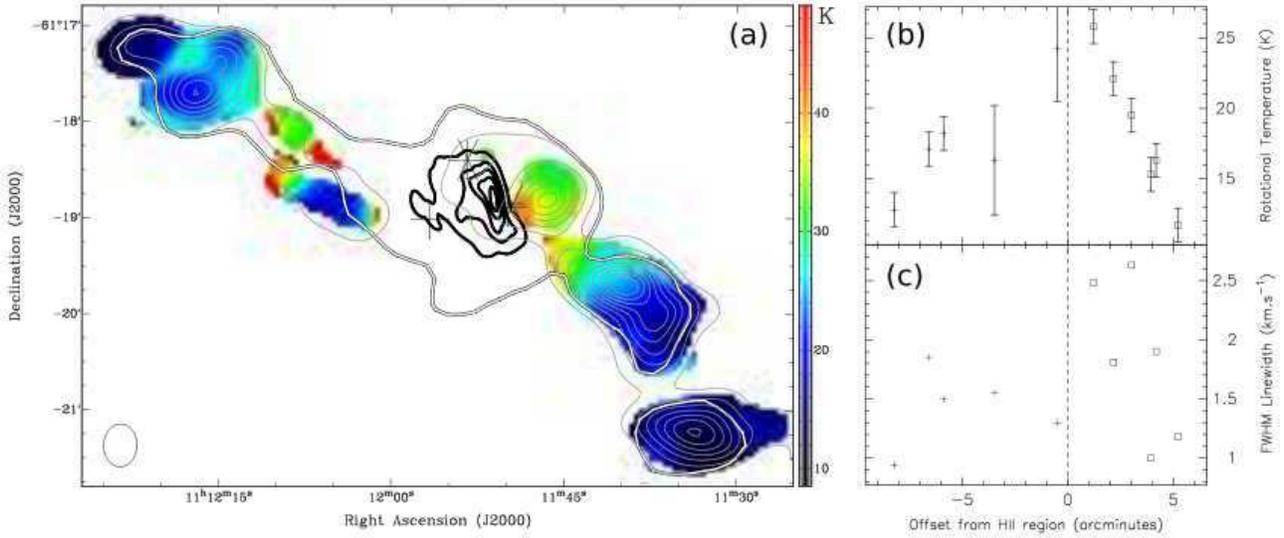}
    \caption[Map of the kinetic temperature in
      NGC\,3576.]{({\bf a}) Map of the kinetic temperature
      derived from the ratio of the \nhthree\,(1,1) and (2,2)
      lines. The beam size is $10.6''\,\times\,11.9''$. Kinetic
      temperatures range from $>$\,40\,K, in the centre of the map, to
      10\,K at the extremes. ({\bf b}) Kinetic temperature and ({\bf c})
    \nhthree\,(1,1) linewidth as a function of angular offset
    from the peak of the H{\scriptsize II} region. Crosses represent
    positions eastwards of the H{\scriptsize II} region and squares
    positions to the west. A distinct temperature gradient is evident in
    the western arm of the filament ({\it top-left}).} 
    \label{fig:nh3temp_dist}
  \end{minipage}
\end{figure*}
Figure~\ref{fig:nh3temp_dist}-a presents a kinetic temperature map of
NGC\,3576 made using the \nhthree\,(1,1) and (2,2) data from the ATCA H75
array observations only. At a resolution of $\sim$\,23\arcsec, the
map reflects the beam averaged temperature of gas above a density of
n\,$\approx\,2\times10^4$\,\cmmthree~\citep{Swade1989}. Higher resolution
observations would doubtlessly reveal regions with hotter or cooler
temperatures than average. An obvious temperature gradient exists
across parts of the filament. Figure~\ref{fig:nh3temp_dist}-b plots
kinetic temperature as a function of angular offset from the peak of
the ionised emission. The data were sampled at 12 positions,
corresponding roughly to the peaks of the \nhthree~clumps in
Figure~\ref{fig:nh311_combined_integ}. The median precision on
the kinetic temperature values is approximately 2.4\,K and stems
from the uncertainties in the Gaussian fits. We find that the kinetic
temperature is higher towards regions adjacent to the H{\scriptsize
  II} region. In the outlying positions the kinetic temperature is
$\sim$\,12\,K, increasing to $\ge$\,30\,K towards the centre. Some hot
spots with temperatures 
above 40\,K exist to the east of the \hii~region (between RAs of
11$^h$12$^m$00$^s$ and 11$^d$12$^m$15$^s$), but the signal-to-noise
ratio in this part of the map is  poor, hence the derived temperatures
are more uncertain ($\sim$\,8\,K). Derived temperatures above
$\sim$\,30\,K have a large associated uncertainty because of
assumptions made during the calculations (see
Appendix~\ref{app:tkin_nh3}). However, below 20\,K \citet{Tafalla2004}
find that kinetic temperatures calculated under the same assumptions
are accurate to better than 5 percent.


\subsubsection{\nhthree~column density and core mass}\label{sec:nh3_column}
Assuming local thermal equilibrium (LTE), the total column density of
\nhthree~may be found directly from the integrated intensity of the
J,K\,=\,(1,1) line via Equation~\ref{eqn:combined_column_total} in the
appendix. An estimate of the total mass of gas
may then be determined from the column density assuming a constant
relative abundance to H$_2$.

\begin{figure*}
  \begin{center}
    \includegraphics[angle=0, width=17cm, trim=0 -20 0 0]{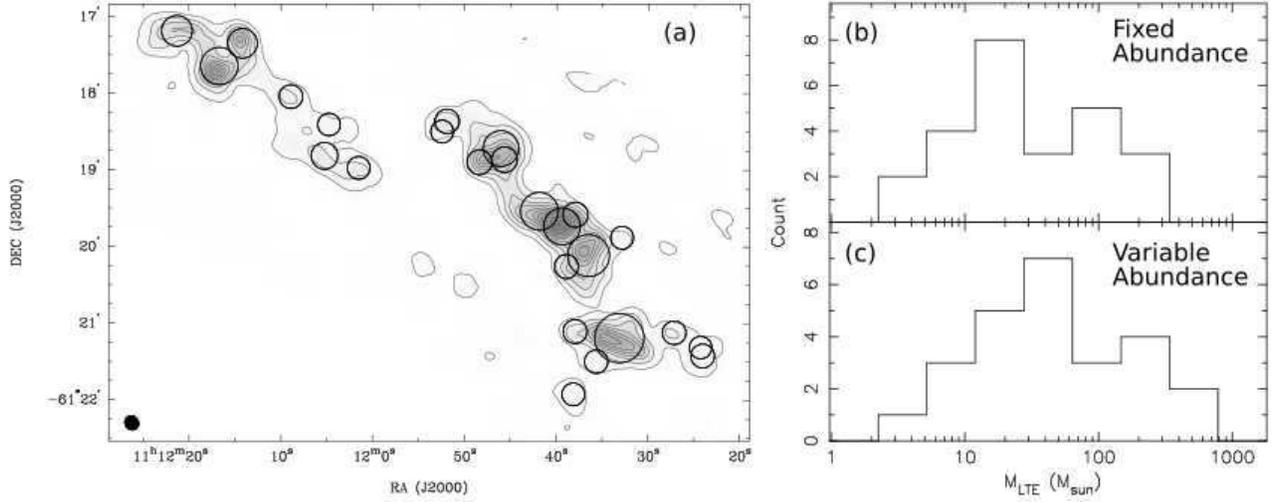}
    \caption[The spatial distribution of cores found in the \nhthree\,(1,1)
      map.]{({\bf a}) The spatial distribution of cores detected in the
      \nhthree\,(1,1) data cube using the {\footnotesize FELLWALKER}
      routine overplotted on the integrated intensity image. Circles
      mark the positions of the cores and the radii are scaled to the
      core mass, as reported in Table~\ref{tab:nh311_clumps}. ({\bf
        b}) Histogram showing the distribution of core masses assuming a single
      value of [\nhthree]\,/\,[H$_2$]=1$\times10^{-8}$. ({\bf c})
      Histogram showing the distribution of core masses corrected for
      the variable abundances calculated in
      Section~\ref{sec:clump_masses}.}
    \label{fig:nh3_clump_positions}
  \end{center}
\end{figure*}

\input tables/11358tb5.tex

In the $\sim$\,11\arcsec~resolution \nhthree\,(1,1) map we begin to
resolve substructure within the 1.2-mm clumps imaged by SIMBA. We have
attempted to decompose the emission using the {\scriptsize
  FELLWALKER}\footnote{The {\tiny FELLWALKER} routine is part of the
  {\tiny STARLINK} package maintained at http://starlink.jach.hawaii.edu/}
routine. {\scriptsize FELLWALKER} attempts to divide regions of emission
into cores by searching for positive gradients in the datacube. The
algorithm considers each pixel in turn above a lower brightness cutoff and
`walks uphill', following the steepest ascent until an isolated local
peak is reached. All pixels visited are assigned to the same core,
which may already exist. Finally, cellular automata fill any holes in
the cores and clean up the edges by replacing each cores'
index with the the most common value occurring within a
3$\times$3$\times$3 pixel cube. The signal/noise
ratio was not constant across the mosaiced \nhthree\,(1,1) map, so the
{\scriptsize FELLWALKER} routine was run on the five SIMBA regions
individually (using the mosaicked data-cube), starting at the same
lower cutoff of 2-$\sigma$, but slightly different tuning
parameters. The results were inspected by 
eye and compared to the original data-cube for consistency. Cores
with fewer pixels than the area of the synthesised beam were omitted
as unreliable detections. 

Twenty-five believable cores were found, whose positions are plotted in
Figure~\ref{fig:nh3_clump_positions}-a. The mass of \nhthree~in each
core was calculated from the sum of the integrated intensities of the
individual pixels in the core. We assumed excitation
temperatures corresponding to those derived from the
23\arcsec~resolution H75 data. The total core mass was then estimated
assuming an abundance ratio [NH$_3$]\,/\,[H$_2$] of 3.0$\times$10$^{-8}$
\citep{Wang2008}, and a correction factor of 1.38 for the abundance of
helium and heavier elements in the interstellar medium
\citep{Allen1973}. Table~\ref{tab:nh311_clumps} 
presents the properties of the detected cores and the distribution of
core masses is illustrated in
Figure~\ref{fig:nh3_clump_positions}-b. In
Section\,\ref{sec:clump_masses}, below, we have calculated the
\nhthree~abundance at different positions in the cloud via a
comparison to 450\,\micron~data. Column~11 presents the core masses
corrected for these results.

Virial masses were calculated from the velocity width and average
radius of the \nhthree~cores reported by {\scriptsize FELLWALKER}.
Neglecting support from magnetic fields or internal heating sources, the
virial mass of a simple spherical system is given by \citep{MacLaren1988}:
\begin{equation}\label{eqn:mvirial}
  {\rm M_{vir}} = k\,{\rm r\,\Delta V^2,}
\end{equation}
where ${\rm M_{vir}}$ is the core mass in \msun, r is the deconvolved
radius of the cloud in parsecs, $\Delta$V is the FWHM velocity width
in \kms~and $k$ is an empirical constant depending on the assumed 
density distribution $\rho({\rm r})$. For $\rho({\rm r})\propto1/{\rm
  r}^{-2}$, $k$\,=\,126 and for $\rho({\rm r})\,=\,$constant,
$k$\,=\,210. Enhanced linewidths (e.g., due to optical depth effects
or blending) will cause us to overestimate the virial mass and the
values quoted here should be considered upper limits. We note
  that $\tau$ varies between 0.1 and 2.0, so line broadening due to
  optical depth effects is negligible. Columns~12 and
13 of Table~\ref{tab:nh311_clumps} present the virial masses derived
assuming constant and 1/r$^2$ density profiles, respectively. 


\subsection{Physical properties from other molecules}
The molecules observed with Mopra towards NGC\,3576 were chosen
specifically to probe the physical conditions in different parts of
the cloud. \thirteenco~(n$_{\rm
  crit}\approx\,6\times\,10^2$\,\cmmthree, see
Table~\ref{tab:transitions3}) and \ceighteeno~emission traces the gas
in the envelope 
surrounding the dense filament. \hcop~and
\hthirteencop~have higher critical densities (n$_{\rm
  crit}\approx\,3\times\,10^5$\,\cmmthree) and so probe conditions at
greater depths, but are also proven tracers of outflows and bulk gas
motions (e.g. \citealt{Rawlings2004}). CS and \ntwohp\, (n$_{\rm
  crit}\approx\,2\times\,10^5$\,\cmmthree) have comparable critical
densities but in practise have been shown to trace dense gas (e.g.,
\citealt{Pirogov2003}, \citealt{Evans1999} and references therein).
\ntwohp~especially, is considered to be 
useful indicator of the coldest ($\leq$\,10\,K) and most dense
regions, where it is predicted to be the most abundant gas-phase ion
\citep{Caselli1995}. This is due to the depletion of its main
destruction partner, CO, onto the dust grains. 

Assuming LTE conditions, we independently solved for the optical depth,
excitation temperature and total column density of CO and \hcop~via
the procedure detailed in the
Appendix~\ref{app:co_col}. Figure~\ref{fig:ngc3576_mopra_columns} 
presents the final column density maps towards NGC\,3576. The
\thirteenco~peak optical depth ranges from 0.1 to 4.4 and the excitation
temperature from 6.4 to 36\,K. Peak \hcop~optical depth ranges from
1.9 to 13.1 and excitation temperatures from 4.2 to 13\,K.
We were unable to independently estimate the excitation temperatures 
of the CS and \ntwohp~transitions and so adopted temperatures derived
from the \nhthree~data. This is a reasonable assumption for \ntwohp~as
we expect both nitrogen bearing compounds to be intermixed in the gas
phase. We were forced to assume that CS was optically thin, which may
not be valid as some of the line profiles appear saturated.

We calculated the total mass of the molecular cloud by assuming the
following abundance ratios to H$_2$: [\thirteenco]\,/\,[H$_2$]\,=\,
6.0$\times\,10^{-6}$ (\citealt{Goldsmith1997}, assuming an
abundance ratio of [$^{12}$CO]\,/\,[H$_2$]\,=\,45 at a galactocentric
radius of 8\,kpc - see \citealt{Langer1990}), [\hcop]\,/\,[H$_2$]\,=\,
2.0$\times\,10^{-9}$ \citep{Zhu2007},
[CS]\,/\,[H$_2$]\,=\,1.1$\times\,10^{-9}$ \citep{Pirogov2007}, and
[\ntwohp]/[H$_2$]\,=\,5.0$\times\,10^{-10}$ \citep{Pirogov2007}. The
values for total cloud mass derived from the CO, \hcop~and
\ntwohp~observations are all approximately 8,300\,\msun. The mass
derived from CS is 12,700\,\msun. These values assume calibration onto
the Mopra `extended beam' temperature scale, which includes power
received from the first sidelobe and is appropriate for data more
extended than 80\arcsec~(see Section~2.1 of \citealt{Ladd2005}).
\begin{figure*}
  \centering
  \includegraphics[angle=00, width=17.0cm, trim=0 -10 0 0]{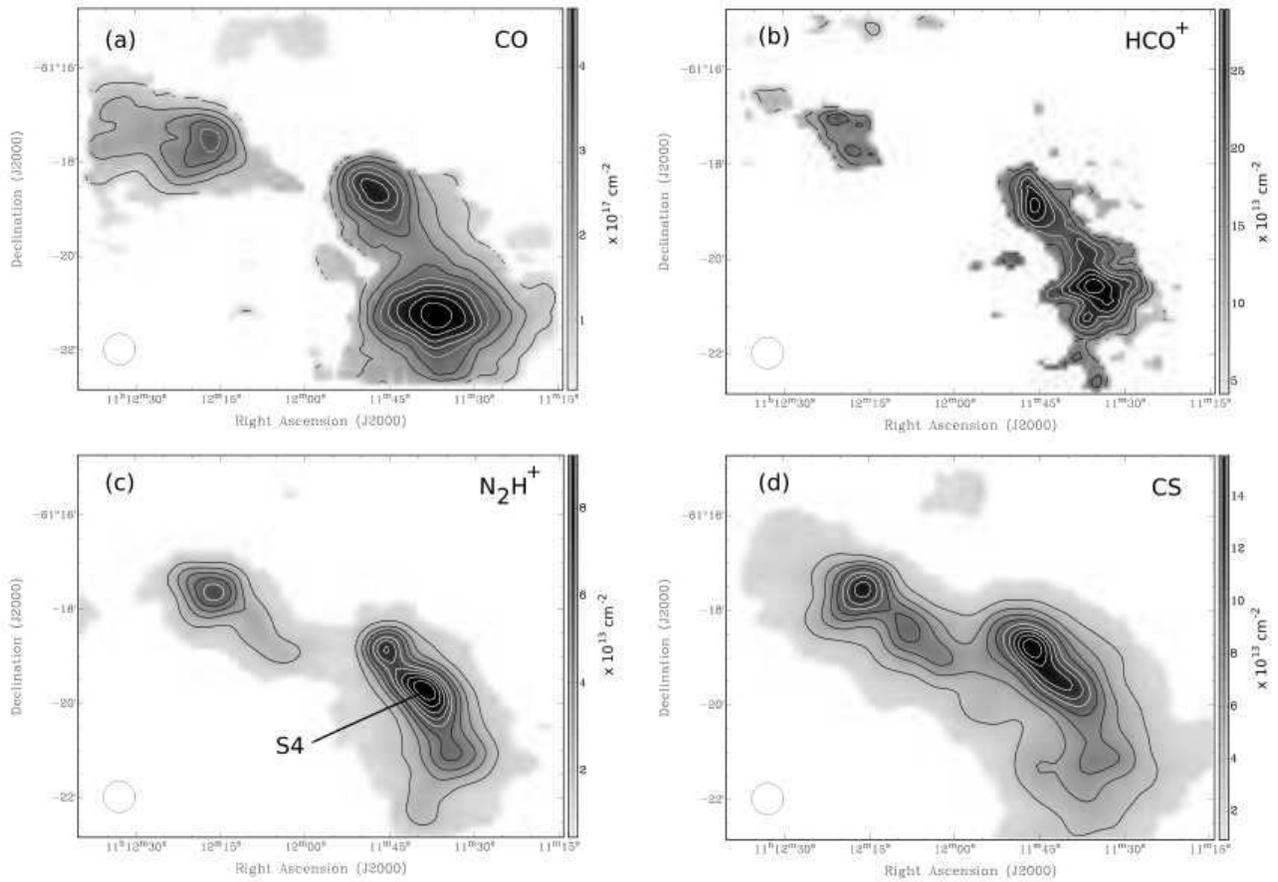}
    \caption{Column density maps of the four species targeted by
      the Mopra telescope. Contours are set at 10 per cent intervals,
      beginning at 10 per cent, except in the \hcop~column density
      map, where they start at 30 per cent of the peak value.}
    \label{fig:ngc3576_mopra_columns}
\end{figure*}


\section{Analysis \& Discussion}

\subsection{\htwoo~masers}
The
three well-known water maser sites within the central \hii~region
(labelled M4, M5 and M9 in Figure~\ref{fig:h2o_masers}) have been
investigated by \citet{Caswell2004}, who summarises their
properties. At the M4 site ({\it G291.274$-$0.709}) the \htwoo~maser
coexists with 6.67\,GHz \chthreeoh~and 1.67\,GHz OH masers. In the
\htwoo~maser spectrum we detect 
bright maser features ranging from $-$37\,\kms~to 23\,\kms, in
agreement with Caswell, however, we also detect weak ($<$\,5\,Jy) lines out to
velocities of $-$51\,\kms~and +20\,\kms. The M5 site ({\it
  G291.270$-$0.719}) hosts a weak \chthreeoh~maser at $-$26\,\kms~and a
bright \htwoo~maser at a velocity of
$-$102\,\kms. First reported by \citet{Caswell1989}, we find this \htwoo~maser
has an unusually broad linewidth (2.9\,\kms) and a peak flux density of
69\,Jy, consistent with the original observations. The V$_{\rm 
  LSR}$ of this spectral feature is reported to have become increasingly
negative since its discovery in 1989 when it had a velocity of
$-$88\,\kms. This has been interpreted as being due to acceleration of
the emitting gas, perhaps in a high velocity outflow
\citep{Caswell2004,Caswell2008}. Alternatively, we suggest that the
$-$102\,\kms~line may be a new maser feature not associated with the component seen at
$-$88\,\kms, which may have been quenched in the intervening time. The
final known site, M9 ({\it G291.284-0.716}), 
exhibits a single intense maser at a velocity of $-$130\,\kms~and with a
linewidth of 4.8\,\kms. We measure an intensity of 670\,Jy in contrast
to the earlier value of 945\,Jy \citep{Caswell1989}, perhaps
reflecting a real decrease in the intensity over the intervening
time. We note that this maser feature falls on the edge of our
bandpass, which may make the flux density measurement unreliable.

The six new maser sites reported here (M1, M2, M3, M6, M7 and M8) are
located in the `arms' of the filament. As can be seen from
Figure~\ref{fig:h2o_masers} all sites are within a few arc-seconds of
\nhthree~emission peaks. With the exception of M8, it is notable
that the spectra of these new masers exhibit, at most, two bright 
features. These features have velocities close to the systemic velocity
and intensities below 10\,Jy. In contrast, the maser spectrum of M8 has
four strong peaks spread over $\sim$\,35\,\kms, the brightest of which
has an intensity of 25\,Jy. The mid-infrared image of the host SIMBA
clump (S5) also exhibits some nebulous emission, on the south-west edge
(see Figure~\ref{fig:summary}-a), while the dense molecular gas tracers (see
Sections~\ref{sec:results_atca}, \ref{sec:results_tid} and
\ref{sec:results_mopra}) show a `notch' in the emission at the same
site. We interpret these features as evidence of a deeply embedded
young stellar cluster evacuating an open-ended cavity.

\subsection{Core masses}\label{sec:clump_masses}
Core masses derived from \nhthree~rely on the assumption of a constant
[\nhthree]\,/\,[H$_2$] abundance ratio, whose value may vary across
the cloud. Independent estimates of the gas + dust-mass in NGC\,3576 have
been made by \citet{Andre2008}, who mapped the 450\,\micron~thermal
emission using the P-ArT\'eMiS\footnote{See
  http://irfu.cea.fr/Sap/en/Phocea/Vie\_des\_labos/Ast/ast\_technique.php?id\_ast=2295}
bolometer camera on APEX\footnote{Atacama Pathfinder Experiment
  http://www.apex-telescope.org/}. With a beam size of 10\arcsec~the
450\,\micron~map has a comparable resolution to our \nhthree\,(1,1)
observations and the morphology of the 450\,\micron~ emission
corresponds almost exactly to the integrated molecular emission.

\input tables/11358tb6.tex
We expanded the analysis of \citet{Andre2008} by decomposing the 2-D
450\,\micron~emission into cores using {\scriptsize FELLWALKER} and
deriving their individual masses via the relation:
\begin{equation}\label{eqn:dust_mass}
{\rm M = \frac{S_{450\,\micron}\,D^2}{\kappa_{450}\,B_{\nu}(T_d)_{450}},}
\end{equation}
where S$_{450}$ is the measured flux
density, $\kappa_{450}$ is the dust opacity per unit (gas + dust) mass
column density at 450\,\micron, B$_{\nu}$(T$_{\rm d}$) is the Planck
function B$_{\nu}$(T) at a dust temperature T$_{\rm d}$ and D is the
distance to the source. We set ${\rm \kappa_{450}=0.04cm^2g^{-1}}$, a value
generally adopted in sub-mm studies of pre-stellar cores
(e.g. \citealt{Motte1998}, \citealt{Ward-Thompson1999}) and
appropriate in regions of moderately high gas densities (around
10$^5$\,cm$^{-3}$ - c.f. \citealt{Henning1995}). For consistency, we
set T$_{\rm d}$ to the kinetic temperatures derived from our
\nhthree~data. A direct comparison between the masses derived from
450\,\micron~and \nhthree~emission integrated over the same apertures
results in estimates of the average [\nhthree]\,/\,[H$_2$] abundance
ratio in each core, assuming no depletion.

Eight equivalent cores were detected in the 450\,\micron~P-Art\'eMiS and
ATCA \nhthree\,(1,1) integrated intensity maps, whose positions
and masses are presented in Table~\ref{tab:clump_mass}. Mass values
from 450\,\micron~data range from $\sim$\,89\msun~to
$\sim$\,712\msun. Some 450\,\micron~cores encompass two or more
\nhthree\,(1,1) cores and in such cases we summed the masses of the
components quoted in column~10 of Table~\ref{tab:nh311_clumps}.
Relative [\nhthree]\,/\,[H$_2$] abundance values
range from 0.4$\times10^{-8}$  
to 7.1$\times10^{-8}$ and are distinctly different in adjacent
cores. We note that the error on the abundance values is
approximately a factor of 3, including the calibration uncertainty of
the P-Art\'eMiS data, assumptions about the dust emissivity and
opacity, dust-to-gas ratio, temperature and relative angular size. The
abundance differences may also be 
attributed to environmental factors. Cores S3-a and -b exhibit the lowest
abundance values, but lie adjacent to the \hii~region. Here they may
be exposed to the ultraviolet-radiation field from the central cluster,
leading to enhanced destruction of large molecules. Core
S4-a, and to a much lesser extent S4-b, have elevated abundances of
\nhthree. Interestingly, this SIMBA clump also has an enhanced
abundance of \ntwohp, possibly due to depletion of CO onto the dust
grain surface (see Section~~\ref{sec:chemistry}, below). In light of
the varying abundance values we have attempted to correct the masses
derived for the {\it cores} found in the 3-D \nhthree~data-cube. Where
  an equivalent 450\,\micron~core does not exist we revert to the
  canonical value of  3.0$\times$10$^{-8}$ \citep{Wang2008}.
 These `corrected' values are presented in column~11 of
 Table~\ref{tab:nh311_clumps}.

\begin{figure}
  \begin{center}
    \includegraphics[angle=0, width=8.5cm, trim=0 0 -20 0]{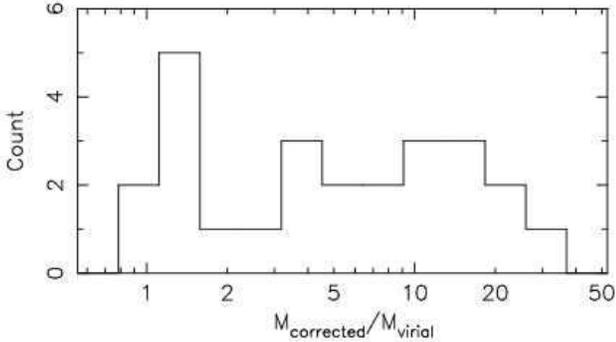}
    \caption{Distributions of M$_{\rm corrected}$/M$_{\rm vir}$ for the 25
      cores found in the \nhthree~data. The M$_{\rm corrected}$ values have
      been derived from the M$_{\rm LTE}$ masses scaled to the
      450\micron~core masses. M$_{\rm vir}$ was calculated assuming a
      assuming a r$^{-2}$ density profile. For a constant density
      profile virial masses would be 1.7 times larger. All
      cores have ratios M$_{\rm corrected}$\,/\,M$_{\rm vir}\,\geq\,1$ and
      are at least consistent with being gravitationally bound. Cores with
      ratios much greater than one are likely to be gravitationally
      unstable and may be undergoing collapse unless supported by
      magnetic fields on the order of milli-Gauss}.
    \label{fig:nh311_mlte_mvir}
  \end{center}
\end{figure}
Figure~\ref{fig:nh311_mlte_mvir} compares the corrected LTE and
virial masses for the \nhthree~cores in
Table~\ref{tab:nh311_clumps}. The factor f\,=\,M$_{\rm
corrected}$\,/\,M$_{\rm vir_1}$ measures the ratio of gravitational to
kinetic energy. Values less than one indicate that the core may be a
transient structure or confined by an over-pressured external medium,
while a value of f\,$\sim$1 suggests that the core is close to
gravitational equilibrium. Cores with f\,$\gg$1 are likely to be
gravitationally unstable, meaning that internal magnetic fields are
required to counteract the effects of self-gravity. We find that in
all cases f\,$\ge$\,1 (within errors) and seventeen cores
have f-values between three and thirty. The magnetic field strength
necessary to support each core may be calculated after
\citet{Bot2007} via
\begin{equation}
{\rm B^2-\,B_0^2 = \frac{9}{10}\left(1\,-\,\frac{10}{9\,f}\right)\,\frac{G\,M^2\,\mu_0}{R^4\,\pi},}
\end{equation}
where B$_0$\,=\,0.3\,nT is the ambient magnetic field strength in the
Milky Way \citep{Han2006}, R is the core radius in metres and M is
the core mass in kilograms. G and $\mu_0$ take their usual values in
SI units. We calculate that field strengths ranging from $\sim$1\,mG
to $\sim$40\,mG, with a median of $\sim$4\,mG, are required to balance
self-gravity in addition to the turbulent pressure. By comparison,
\citet{Curran2007} conducted polarimetry measurements of the magnetic
field strengths in fourteen massive star forming regions and found
values from $<$\,0.1\,mG to $\sim$6\,mG. Our values are higher on
average, suggesting that some of our \nhthree~cores may indeed be
collapsing. In Section~\ref{sec:kinematics} below we present evidence
of collapse in the SIMBA clump S4.

\citet{Andre2008} estimated the envelope mass of the {\it dominant}
protostellar object in each 450\micron~core by scaling
the mass corresponding to the peak flux density to a uniform diameter
of 6,000\,AU, assuming a density profile $\rho({\rm r})\propto$\,r$^{-2}$. The
values range from 21 to 45\,\msun. They also estimated the bolometric
luminosities by fitting the SIMBA, P-ArT\'eMiS and
MSX\footnote{Midcourse Space Experiment (MSX)\\
  ~~~http://irsa.ipac.caltech.edu/applications/MSX/} data with a grid of
spectral energy distribution models computed by \citet{Robitaille2006,
  Robitaille2007}. When plotted on a  M$_{\rm env}$ versus  L$_{\rm
  bol}$ diagram (see Figure~4 of \citealt{Andre2008}) the objects are
bracketed by evolutionary tracks corresponding to final stellar masses
of 15 to 50\,\msun, implying that the cores are undergoing
high-mass star-formation. The relative positions of the cores on the
diagram also indicates their relative evolutionary
ages. Interestingly, the cores furthest from the \hii~region, at
positions S1-b, S1-c, S4-a, and S5, are significantly less evolved
than the adjacent cores at S3-a and S3-b. This evolutionary
gradient is consistent with the hypothesis that the central
\hii~region has triggered sequential star-formation in the filament.


\subsection{Chemistry}\label{sec:chemistry}
\begin{figure*}
  \begin{center}
    \includegraphics[width=17cm, trim=0 0 0 0]{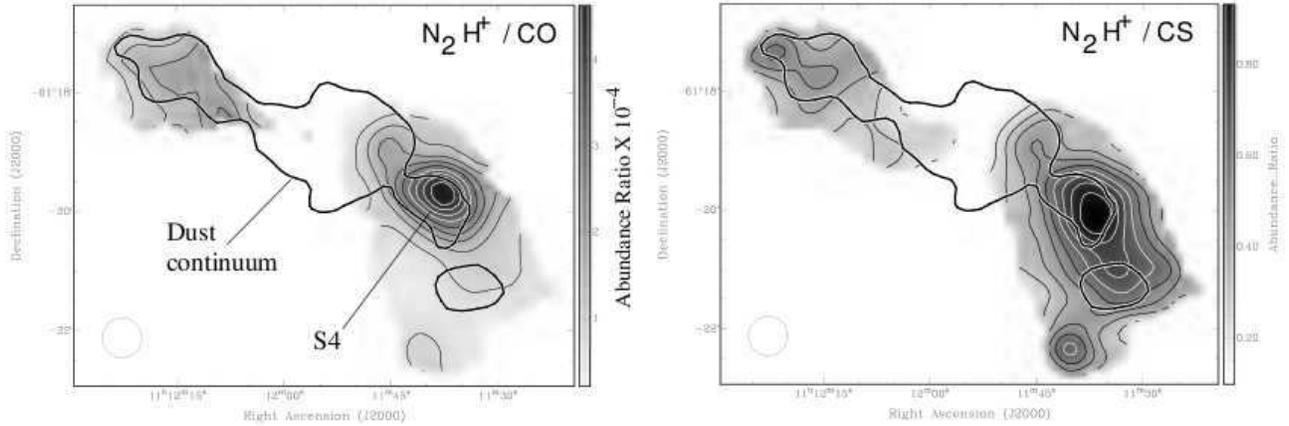}
    \caption[Maps of the relative abundance of \ntwohp~vs CO and
      CS.]{Maps of the relative abundance of [\ntwohp]\,/\,[CO] 
      ({\it left}) and [\ntwohp]\,/\,[CS] ({\it right}) made from individual
      column density maps. Greyscale and contours are set at 10 per cent
      intervals, starting at 30 per cent of peak. The
      \ntwohp~abundance is significantly enhanced towards clump
      S4. This difference is especially pronounced in the [\ntwohp]\,/\,[CO]
      map but is not as clear when comparing \ntwohp~and CS. We note
      that the column density of CS is the least well known of the
      three species.} 
    \label{fig:rel_abundances}
  \end{center}
\end{figure*}
Once the physical conditions of temperature, density and optical-depth
have been determined, we can begin to examine the chemistry in
NGC\,3576. Ratios of column density provide beam-averaged measures of
the relative abundances between two species. For key molecules this
ratio is tied directly to the physical conditions and hence to the
embedded star formation activity.

Figure~\ref{fig:rel_abundances} presents maps of the relative
abundance of [\ntwohp]\,/\,[CO] and [\ntwohp]\,/\,[CS] made by taking
the ratios 
of the column density maps. The abundance ratios are beam-averaged and
assumes all of the gas is above the critical density for each species.
Immediately clump S4  stands out in the [\ntwohp]\,/\,[CO] map, as it has an
over-abundance of \ntwohp~compared to the other SIMBA clumps. The
difference is not as pronounced in the [\ntwohp]\,/\,[CS] map,
however, we note the optical depth of CS is not known and hence there
are significant uncertainties in the column density across the map. A
comparison between the \ntwohp~and \hcop~column densities
also reveals the same enhancement in clump S4, which we
believe to be a real chemical difference. Because of the low
signal-to-noise we do not show this map.

The abundance of \ntwohp~is predicted to be enhanced in cold
($\leq$\,10\,K) or dense clumps. This is because its parent molecule
  N$_2$ is one of the least affected by the condensation process and
  \ntwohp~survives in the gas phase, at least for densities in the
  range 10$^5$\,--\,10$^6$\,\,\cmmthree. In such environments its main
destruction partner CO depletes onto the grains and the abundance of
\ntwohp~consequently increases \citep{Bergin2002}. Interestingly,
\thirteenco~and \ceighteeno~emission are noticeably absent from clump
S4, providing further confidence in this interpretation. Conditions
such as these are found in starless cores, but particularly in cold
collapsing clumps evolving towards forming stars. Clump S4 has an
average kinetic temperature of $\sim$\,25\,K according to our analysis
of \nhthree~data. This seems high for an infrared dark cloud (see
  Figure~\ref{fig:summary}-a) undergoing collapse, however, CO
depletion may still occur at these temperatures if the density is high
enough \citep{Aikawa2001}. In the following sub-section we will
examine the evidence for inflowing gas motions in this clump. 


\subsection{Morphology and kinematics}\label{sec:kinematics}
\begin{figure*}
  \centering
    \includegraphics[width=12cm, trim=0 0 0 0]{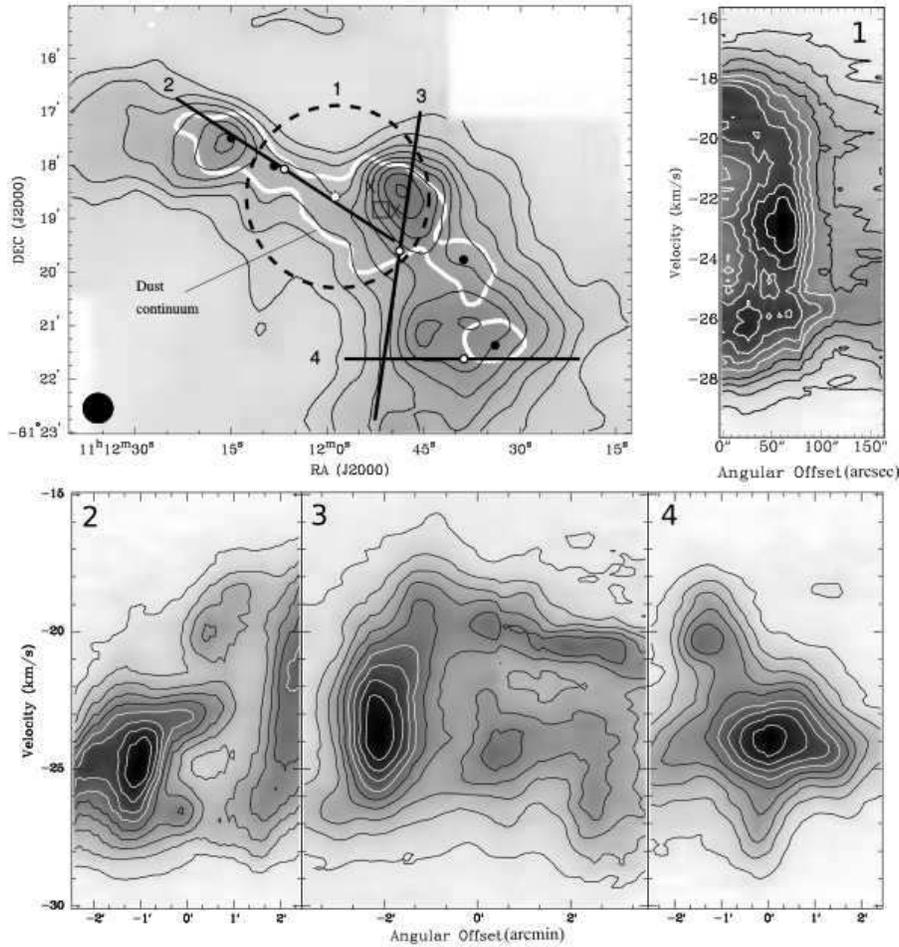}
    \caption[Position-velocity cuts through the
    \thirteenco~data-cube.]{~Position-velocity cuts through the
    \thirteenco~data-cube. The top-left panel is an integrated
    intensity map of \thirteenco~showing the orientations and
    positions of the PV-slices presented in panels 1\,--\,4.
    Centre positions of the slices are marked with black and white
    filled circles.
    Panel 1 shows the azimuthly averaged intensity as a function of
    velocity and angular offset from position 11$^h$11$^m$58.0$^s$
    $-$61$^d$18$^m$35.0$^s$\,J2000. The `C' shaped profile is
    indicative of an expanding shell of emitting gas. Panels 2, 3 and
    4 present PV-slices along the axes of emission and in
    cross-section through the western arm, respectively. Several distinct
    filamentary structures are evident, merging into the bright
    emission adjacent to the \hii~region.}
    \label{fig:ngc_13co_pv}
\end{figure*}
We have examined the molecular emission for evidence of velocity
gradients indicating bulk gas motions, expansion or
contraction. Figure~\ref{fig:ngc_13co_pv}~presents position-velocity
(PV) diagrams made using the \thirteenco~Mopra data and annotated on
the inset integrated intensity map. Panel 1 was made using
the {\it kshell} tool\footnote{{\it kshell} is part of the KARMA suite
of analysis software available from the ATNF at
http://www.atnf.csiro.au/computing/software/.} and shows the azimuthly 
averaged intensity  as a function of velocity and angular offset from
position 11$^h$11$^m$58.0$^s$ $-$61$^d$18$^m$35.0$^s$\, J2000. The
`C' shaped profile is a classic indicator of an expanding shell of
gas, in this case centred approximately on the \hii~region. Towards
the centre position (zero offset) the spatial pixels sample
gas moving directly towards and away from the observer, hence the
average spectrum peaks at the systemic velocity $\pm$ the shell
expansion velocity. At offsets approaching the shell radius the pixels
sample gas moving perpendicular to the observer, hence the emission
peaks at the systemic velocity ($-$24\kms). The clumpy shell seen here
in \thirteenco~is approximately 1 arcminute in diameter and
encompasses the free-free emission from the \hii~region. We see
exactly the same C-shape in \ceighteeno, although the plot is
noisier.

Panels 2\,-\,4 present the PV-slices indicated on the inset
\thirteenco~moment map (top-left). The first cut, (centred at 11$^h$11$^m$09.90$^s$,
$-$61$^d$15$^m$45.6$^s$ J2000 and at a position angle $\Theta$\,=\,244.5\degrees)
is a longitudinal slice through the eastern arm of the filament and
reveals a forked cross-section: two tenuous tentacles of emission 
merge into the eastern-most SIMBA clump (S1). PV-slices through the
dense gas tracers (\nhthree~and CS) show the same structure, meaning
that this morphology is not due to optical depth or chemical
effects. Ionised gas from the \hii~region overlays this part of the
eastern arm, likely inter-penetrating the molecular emission. Combined
with the existence of an expanding shell and a temperature gradient, we
conclude that the \hii~region is expanding eastwards, and has played a
significant role in sculpting these features.

The second cut (11$^h$11$^m$48.60$^s$, $-$61$^d$20$^m$57.6$^s$ J2000,
$\Theta$\,=\,174\degrees) follows the line of emission evident in
Figure~\ref{fig:ngc_13co_pv} at $-$20\,\kms. It is clear from the
PV-diagram that two velocity components exist, separated by
$\sim$\,4\,\kms. Both appear to merge into the strong emission
immediately west of the \hii~region. These two velocity features are
clearly visible in the third cut (11$^h$11$^m$39.50$^s$,
$-$61$^h$21$^m$46.9$^s$ J2000,
$\Theta$\,=\,268\degrees), which presents the velocity structure of the
western arm in cross-section. The bulk of the emission in this arm
derives from the component at $-$24\,\kms, while the component at
$-$20\,\kms~ is not seen in the dense gas tracers \nhthree, \ntwohp~or
CS. Considering the dense gas only, we see the peak V$_{\rm lsr}$ of
the filament is approximately constant between $-$24 to
$-$25\,\kms~over its length. No large opposing velocity gradients or
discontinuities are evident and small variations may be explained by
optical depth effects or motions on small scales
($<$\,30\arcsec). We see no evidence for large scale flows of 
gas, at least in the 40\arcsec~resolution Mopra data, and conclude that the
majority of the emission derives from the same cloud and is not merely
a projection or line-of-sight effect. The velocity structure in the
11\arcsec~resolution \nhthree~data is complex within the bounds of
each SIMBA clump and will be analysed in detail in a separate paper.

\begin{figure*}
  \centering
    \includegraphics[angle=0, width=15.5cm, trim=0 0 0 0]{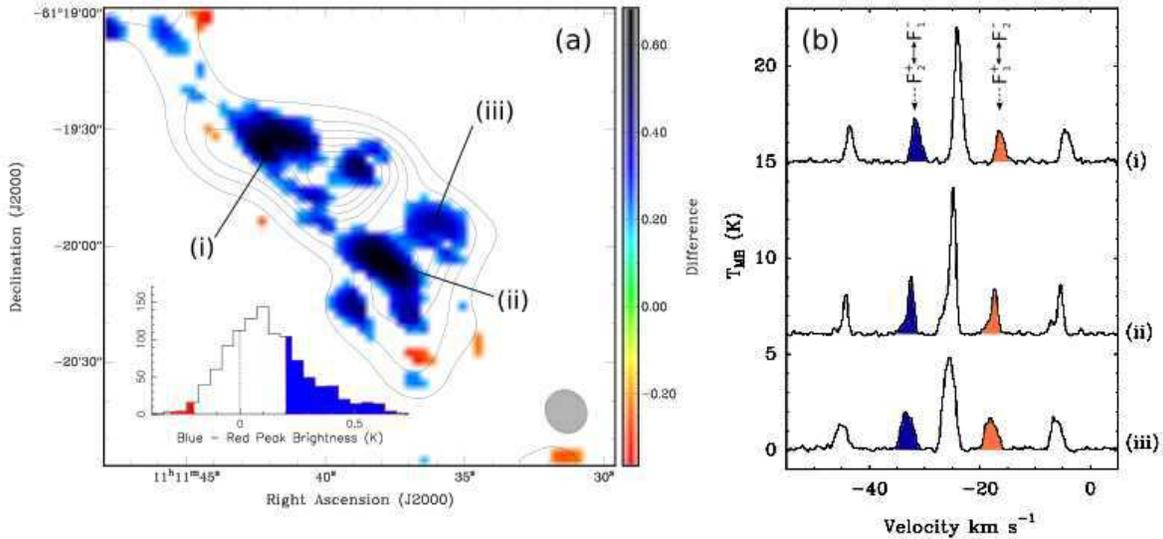}
    \caption{({\bf a})  Map of the difference between the peak
      intensities of the F\,$^+_2\leftrightarrow\,$F\,$^-_1$ (blue) minus
      F\,$^+_1\leftrightarrow\,$F\,$^-_2$ (red) {\it inner} satellite lines
      of \nhthree\,(1,1) for SIMBA clump S4, where F is the total
      angular momentum quantum number (see Figure~1 in
      \citealt{Rydbeck1977}).} Only spectra with 
      differences greater than 0.2\,K ($\sim\,$3-$\sigma$) are plotted
      as colourscale, while contours represent the integrated
      intensity map. The inset histogram shows the distribution of
      difference values over the whole of the clump and is clearly
      dominated by significant numbers of blue-skewed spectra. ({\bf
      b}) Sample spectra from the positions marked i, ii and iii on
      the difference map.
  \label{fig:S4_cgspec}
\end{figure*}

We are interested in probing for bulk gas motions in clump S4, where we
find an over-abundance of \ntwohp, common in collapsing
clumps. Asymmetric spectra towards 
star forming regions are often interpreted as indicators of inward or
outward motions. In the case of \nhthree\,(1,1) asymmetries may be
produced by a combination of non-LTE effects and bulk
motions. \citet{Park2001} has modelled \nhthree\,(1,1) spectra over a
wide range of physical conditions, predicting that asymmetries
between the {\it inner} satellite lines indicates inward or outward
flows of gas. On the other hand, asymmetries between the outer
satellites are indicative of selective radiative trapping and non-LTE
excitation \citep{Stutzki1985}. The amount by which both effects skew
the profiles increases with optical depth. Figure~\ref{fig:S4_cgspec}-a
presents a map showing the difference between the blue minus red peak
intensities of the {\it inner} satellite \nhthree\,(1,1) lines, for
SIMBA clump S4. Only pixels with absolute differences greater than 0.2\,K
($\sim$\,3-$\sigma$ above the spectral noise) have been
plotted. Asymmetries on the order of $\sim$\,10 percent are observed
over distinct regions within the SIMBA clump. We note that there are
few pixels with significant red-skewed spectra (values less than
$-$0.2\,K in Figure~\ref{fig:S4_cgspec}-a) and the map is dominated by
regions of blue-skewed spectra which cover angular areas greater than
the beam. Sample spectra from the three largest regions are presented
in Figure~\ref{fig:S4_cgspec}-b. The optical depth of \nhthree\,(1,1)
in this clump ranges between $0.4\leq\tau\leq1.0$ and we do not expect
large asymmetries under these conditions, hence, we believe the
differences are significant and tentatively imply inward gas motions.

\subsection{\hii~region -- cloud interaction}
It is clear from the temperature map presented in
Figure~\ref{fig:nh3temp_dist} that the central star-forming complex is
heating the gas in its immediate surroundings. The heating is gradual
on the western side of the filament and the temperature decreases
linearly towards the extremes. 
The gradient in temperature strongly suggests that the 
H{\scriptsize II} region is embedded within the filament and is not
merely a line of sight projection. The temperature fluctuates between
$\sim$\,15 and $\sim$\,30\,K in the S2 clump, immediately east of the
\hii~region's peak. Diffuse free-free emission is observed to extend
into this region and the molecular gas has a clumpy distribution and
is less evenly heated. It is interesting to note that there is a weak
linewidth gradient in the western arm also, as plotted in
{Figure~\ref{fig:nh3temp_dist}-c} . This may reflect the turbulent
energy injected into the cloud via shocks driven by the expanding
ionised gas in the \hii~region. Clear evidence for this continuing
expansion is seen in the \thirteenco~data as an expanding shell of
molecular gas surrounding the free-free emission.


\subsection{Infrared excess stars}
\begin{figure}
  \begin{center}
    \includegraphics[angle=0, width=8.5cm, trim=0 0 0 0]{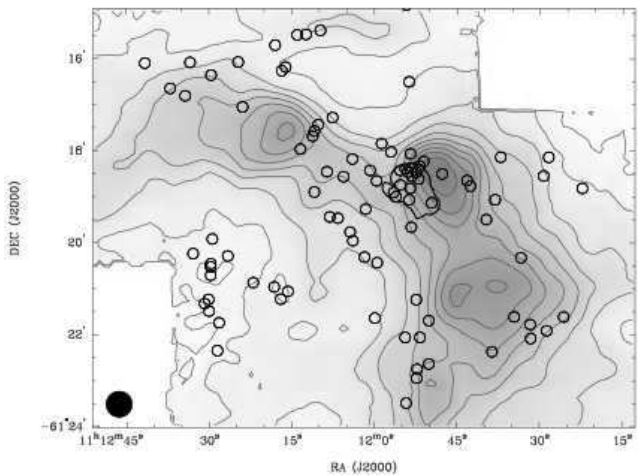}
    \caption[The spacial distribution of stars with infrared
      excess.]{~The distribution of stars with infrared excess
      determined by \citet{Maercker2006}, plotted  over the molecular
      gas traced by \thirteenco\,(1\,--\,0) emission.} 
    \label{fig:ir_excess}
  \end{center}
\end{figure}
Previous studies have focused on the infrared properties
of embedded stars in NGC\,3576. Emission in the K (2.2\microns) and L
(3.6\microns) bands, which cannot be accounted for by interstellar
reddening, has been interpreted as the signature of thermal emission
from a dusty disk. \citet{Maercker2006} imaged the region in J
(1.3\microns), H (1.7\microns), K and L, and found 113 stars with
infrared excess. Their distribution is plotted in
Figure~\ref{fig:ir_excess} over the \thirteenco\,(1\,--\,0) integrated
intensity map. Foreground stars have been filtered from the list. A
dense cluster of IR-excess stars is centred on the Giant HII region,
as reported in work by \citet{Persi1994} using Las Campanas
Observatory. It is clear that the distribution of IR-excess stars is
anti-correlated with the  molecular gas in the dense arms of the
filament. The high extinction in this area likely makes it impossible
to detect even bright IR-excess sources behind or embedded in the
filament. This suggests that the clumps are not associated with
star-formation at an {\it advanced} stage of development. If 
young stars are present in the filament, they are probably at such a
young age that they are still deeply embedded and not yet visible in
the near-infrared. 


\subsection{Triggered star formation in NGC\,3576?}
\begin{figure*}
  \begin{center}
    \includegraphics[angle=0, width=18.0cm, trim=0 0 0 0]{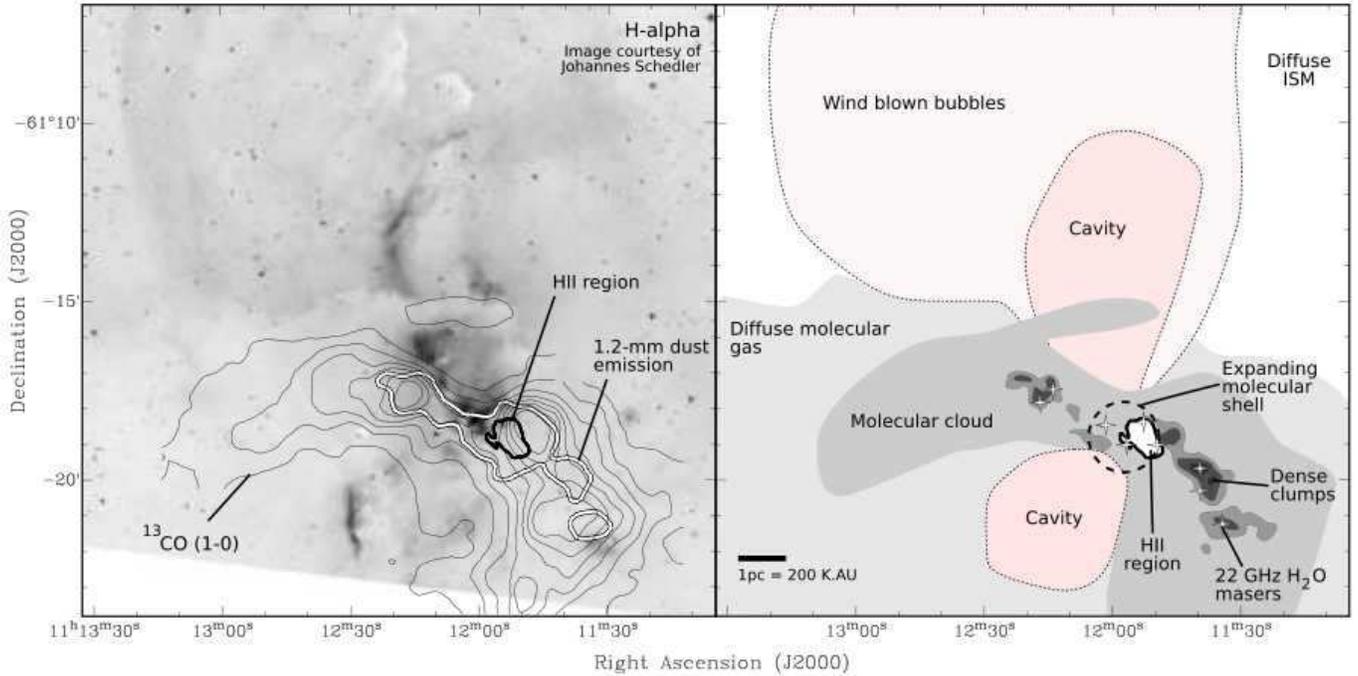}
    \caption{~Schematic of the environment of
      NGC\,3576 ({\it right panel}). \thirteenco~integrated intensity
      contours tracing the extended molecular cloud, overlaid on a
      H-$\alpha$ image ({\it left panel}), which highlights the large
      scale structure of the interstellar medium around the
      star-forming complex. The molecular emission is correlated
      exactly with the extincted regions of the H-$\alpha$. Loops of
      H-$\alpha$ emission are seen to extend north and south from the
      vicinity of the ionised \hii~region (thick black contour),
      likely shaped by the energetic young cluster at the heart of the
      complex. The ionised gas is expanding into the molecular cloud,
      sweeping up a clumpy shell of gas and heating the filamentary
      cloud in which it is embedded. Dense cores containing tens to
      hundreds of solar masses have formed along the length of the
      cloud and exhibit 22\,GHz water maser emission, signifying
      star-formation in progress.} 
    \label{fig:cartoon}
  \end{center}
\end{figure*}
Our observations provide an interesting glimpse into star formation
activity within NGC\,3576. It is clear that the \hii~region is
embedded within the dusty filament and is exerting a powerful
influence on its immediate surroundings. We have investigated whether
the \hii~region expansion into the ambient molecular cloud is
consistent with triggering the formation of high-mass stars
along the dusty filament.

By comparing the morphology of the \hii~region (traced by the 23\,GHz
continuum emission) to the dense molecular gas (traced by \nhthree, CS
and \ntwohp) and thermal dust emission (traced by 1.2-mm and
450\micron~continuum), we see that the young star cluster at the heart
of the complex has destroyed or dispersed much of the gas in the
central clump. There is also clear evidence for an expanding shell in
the \thirteenco~and \ceighteeno~data, centred on the \hii~region.

To the east of the  \hii~region, molecular gas is sparse and does not
follow the morphology of the dust, while at 3.4-cm wavelengths the ionised gas is
observed to extend throughout most of field S2 (see
Figure~\ref{fig:summary}-c). In contrast, the western edge of the
free-free continuum emission is observed to have a well defined
boundary. We find from our molecular-line observations that this
edge coincides with a dense molecular clump, implying that the
\hii~region is pressure-confined on this side. At the same time, the
\hii~region appears to be heating the arms of the filament and the
western arm exhibits a clear temperature gradient, peaking adjacent to
the \hii~region. In the eastern arm, the clumpy nature of the gas gives
rise to `hot spots' with higher temperatures, consistent with our
`dispersal' interpretation.  The gradients in temperature strongly
suggest that the \hii~region is embedded within the filament.
Some of the structures visible in the eastern arm are reminiscent of
the `elephant's trunks' pillars in the Eagle Nebula \citep{White1999}. 

Besides the heating evidence, there is also evidence for kinematical interaction 
between the  \hii~region and the ambient molecular cloud. We
do see evidence that the \hii~region is expanding 
into the eastern arm, sweeping up a clumpy shell of gas. Such an expansion is
expected in the collect and collapse scenario for triggered star formation (see Section~\ref{sec:intro}).
Multiple velocity components have been identified within the arms
of the filament (see Figure~\ref{fig:ngc_13co_pv}). Velocity gradients
and dislocations may be the archaeological remains of cloud-cloud
collisions or quickly collapsing elongated clumps
(c.f. \citealt{Peretto2006}).  However, when considering only the
dense gas we see that the velocity structure of the gas is smooth, 
confined to a narrow range and does not support either
interpretation, at least at the 40\arcsec scale of the Mopra data. 

Different diagnostics have been employed to identify signatures of early phases of star formation in the
filaments and gas clumps. Water masers have always been associated with star formation (see
\citealt{Beuther2002} for a summary). That we
detect water-maser emission towards the arms of the filament confirms
the presence of shocks and hence that stars are forming within the
molecular gas. However, unlike 6.67\,GHz \chthreeoh~masers, 22\,GHz
water masers are not a clear 
confirmation of high-mass star formation as they are
also found towards sites where intermediate mass (4\,--\,8\msun) stars
are forming. Water masers are thought to originate in the shocked gas
associated with outflows and are collisionally pumped (e.g.,
\citealt{Kylafis1992}). 
Core masses derived from \nhthree~range from
5\,\msun~to $\sim$\,516\,\msun, with a median value of
$\sim$\,45\,\msun. Seven cores have masses greater than 100\,\msun~and
for a typical star formation efficiency of 0.3
(c.f. \citealt{LadaLada2003}) are likely to form stars with masses
greater than 8\,\msun.

We have confirmed that protostellar objects are likely forming in the arms of
the filament and that the \hii~region is clearly influencing the
conditions in the bulk of the cloud, as evidenced by the temperature
gradients and gas-morphology. Whether the formation of the new stars
is triggered by the expansion of the \hii~region is difficult to prove
categorically, but to be consistent requires evidence of an `age
gradient', as argued by \citet{Persi1994}.  Recent 
  450-$\mu$m continuum observations \citep{Andre2008} partially
  resolved the 1.2-mm continuum SIMBA clumps and detected continuum
  emission counterparts to the NH$_3$ cores.  Based on a `protostellar
  mass envelope vs. bolometric 
  luminosity diagram' (hereafter $M_{env}$\,--\,$L_{bol}$ diagram),
  \citet{Andre2008} estimated the evolutionary stage of these candidate
  massive protostars and positioned them on evolutionary tracks for
  high mass star formation. Interestingly, S3-a and S3-b, the
  closest candidate protostars to the \hii~region, appear to be the most evolved
  objects on the $M_{env}$\,--\,$L_{bol}$ diagram. In contrast, the
  other candidate protostars, those in the arms, appear to be less
  evolved on the diagram and are possibly class 0-type
  protostars. The non detection of ionised gas toward these protostars
  confirms their very early evolutionary stage. 
  All detected protostars will evolve into stars of masses between
  15 and 50 M$_{\odot}$ according to the evolutionary tracks on the
  $M_{env}$\,--\,$L_{bol}$ diagram. 

Another signature of early phases of star formation is the evidence for collapsing
cores. On examination of the relative molecular abundances, SIMBA clump S4
stands out as having an over-abundance of \ntwohp~and an
under-abundance of CO. Such a chemical signature is typical of dense
and cold regions, which is often found towards collapsing cores. We
searched for signatures of inflowing motions in the \nhthree~line
profiles. Asymmetries on the order of $\sim$\,10 per cent are observed
between the inner satellites for a region running lengthwise along the
clump S4. 

High-mass star formation is clearly underway in the dusty arms of
the NGC\,3576 filament. Evidence for interaction through temperature
gradients as well as an evolutionary sequence for the
embedded protostars from the \hii~region to the arms, is consistent
with the hypothesis that the \hii~region is triggering star
formation in NGC\,3576. The geometry of the \hii~region
and associated clumps does not resemble the typical collect and
collapse scenario, hence the observations are also in agreement with
spontaneous star formation, at least in the clumps further away from
the ionisation front. However, it is also possible that the
\hii~region is responsible for initiating star formation in these
clumps via shock driven globule squeezing. Further investigation is
necessary to distinguish between these two cases.


\section{Summary \& Conclusions}
We have mapped the molecular environment of the giant \hii~region NGC\,3576
in lines of \nhthree, CO, \hcop, CS and \ntwohp. In addition, we have
searched for \uchii~regions via 23\,GHz free-free emission and
for 22\,GHz water masers. Figure~\ref{fig:cartoon} summarises our
observational results as a schematic, overlaid on an amateur
H-$\alpha$ image (J. Schedler, 2009) which illustrates the large
scale structure of the ISM in the region. Our main conclusions are as
follows: 
\begin{enumerate}
  \item We have detected molecular emission from all clumps
  identified in the 1.2-mm SIMBA map \citep{Hill2005}. Dense gas
  tracers (\nhthree, 
  \ntwohp~and CS) follow the morphology of the thermal dust emission,
  except within the \hii~region, where the 1.2-mm emission is
  highly contaminated by free-free continuum emission. Molecules with
  low and intermediate critical densities or 
  with high abundances (CO, \hcop) trace the extended envelope well.
  \item We searched for \uchii~regions in the arms of the filament
  via their 23\,GHz emission, but found none down to a detection limit
  of 0.5\,mJy/beam. The central \hii~region
  was mapped and exhibited a similar morphology to the earlier
  3.4-cm image by \citet{DePree1999}. 
  \item Six new sites of 22\,GHz water maser emission were detected,
  all of which lie adjacent to discrete \nhthree~cores in the arms of the
  filament. Most of the new masers have flux densities less than ten
  Janskys, compared to the three known sites in the \hii~region, which
  have flux densities between 40\,--\,640\,Jy. Of all the new masers,
  the site in clump S5 has the greatest range of velocities and
  warrants further investigation.
  \item \nhthree\,(1,1) and (2,2) were used to derive the kinetic
  temperature of the dense gas. We find a temperature gradient exists
  in the western arm, with temperatures ranging from $\sim$\,11\,K at
  the extremes to $\ge$\,30\,K adjacent to the \hii~region. Clearly, the
  \hii~region is responsible for the heating.
  \item There is clear evidence in \thirteenco~and
    \ceighteeno~for an expanding swept-up shell of diameter
    $\sim$\,1\arcmin~completely encompassing the \hii~region.
  \item The $\sim$\,11\arcsec~resolution \nhthree\,(1,1) emission
  was decomposed into twenty-five individual cores, ranging in mass
  from 5\,\msun~to 516\,\msun, when scaled to the 2-D clump masses
  derived from the 450\,\micron~P-ArT\'eMiS map. 
  \item  M$_{\rm corrected}$\,/\,M$_{\rm vir}\ge$\,1 for all
  \nhthree~cores, implying they are at least
  gravitationally bound. Seveneen cores have M$_{\rm
  corrected}$\,/\,M$_{\rm vir}\ge$\,3 and to avoid 
  collapsing under self-gravity require support from internal magnetic
  fields with strengths between $\sim$1\,mG and
  $\sim$\,40\,mG. These values are on average higher than measured in
  recent observations of massive star-formation regions and suggest
  that some of the cores may be collapsing.
  \item Clump S4 in Figure~\ref{fig:summary} is over-abundant in
  \ntwohp~and shows evidence for depletion of CO, consistent with the
  expected chemistry in a cold, dense clump. We examined our~\nhthree~data for
  line-profile asymmetries indicating bulk gas motions and found
  distinct regions with blue profiles, tentatively indicating inflowing gas.
  \item The filament displays a complex velocity structure, with two
  main components in each arm seen in the \thirteenco~data. Molecules
  tracing high density gas show that a single component at
  V$_{\rm LSR}$\,=\,$-$24\,\kms~dominates the cloud. 
  \item The positions of stars with infrared-excess, indicative of
  circumstellar disks, are anti-correlated with the dense molecular
  gas, implying star formation has not yet advanced far enough to be
  visible in the near-infrared. Alternatively, there is likely
  sufficient extinction to hide such stars embedded within the
  molecular cloud.
\end{enumerate}
  
From the above evidence it is clear that star formation is
underway across the whole of the NGC\,3576 filament and that the
\hii~region is influencing the physical conditions in the bulk of
the molecular gas. Very young massive protostellar objects,
some equivalent to low-mass class-0 protostars, are forming within
the arms of the filament. Masses and luminosities of cores with
1.2-mm, 450\micron~and infrared data have been
characterised by \citet{Andre2008}. Objects adjacent to the
\hii~region appear to be more evolved, implying an evolutionary
gradient. The \hii~region is expanding into the cloud, dissipating the gas
on its eastern edge and is possibly the trigger for
  star-formation in the nearest clumps. However, the observational
  data is also consistent with spontaneous star formation, at least in
  the outlying regions.


\section*{Acknowledgements}
The Mopra radio telescope is part of the Australia Telescope which is
funded by the Commonwealth of Australia for operations as a National
Facility managed by CSIRO. During 2002\,--\,2005 the Mopra telescope
was operated through a collaborative arrangement between the
University of New South Wales and the CSIRO. 

We wish to thank the Australian research council and UNSW for grant
support. CRP was supported by a School of Physics Scholarship during
the course of his PhD.

V. Minier and F. Herpin acknowledge the use of a French-Australian
Science \& Technology (FAST) Program grant. The FAST program is jointly
managed by the Department of Innovation, Industry, Science \& Research
and its French counterparts, the Ministry of Higher Education and
Research and the Ministry of Foreign and European Affairs. 

PAJ acknowledges partial support from Centro de Astrof\'\i sica FONDAP
15010003 and the GEMINI-CONICYT FUND.

Many thanks also to Johannes Schedler, who kindly provided his
H-$\alpha$ image of NGC\,3576.

We are also grateful to the anonymous referee for very thorough
comments and discussions that helped improve the presentation.


\bibliography{11358}


\appendix
\section{Physical properties from molecular lines}
This work makes heavy use of standard methods to derive physical
properties from molecular line data. Here we summarise the methodology
used and gather the equations and constants in one place. All
calculations were performed in SI units. 

\subsection{Kinetic temperature from \nhthree}\label{app:tkin_nh3}
Firstly, we calculate the optical depth of the
\nhthree\,(1,1) transition via a comparison of brightness temperature
T$_{\rm B}$ in the main and satellite lines: 
\begin{equation}
  {\rm \frac{T_{B,m}}{T_{B,s}}=\frac{1-e^{-\tau_m}}{1-e^{-a\tau_m}}.}
\end{equation}
The subscripts `m' and `s' indicate the main and satellite
groups, respectively, and `a' is the theoretical T$_{\rm B,s}$/T$_{\rm B,m}$
ratio under optically thin conditions. For \nhthree(1,1), the
  fractional intensity in the main group is 0.502 and in individual
  inner and outer satellite lines is 0.139 and 0.111,
  respectively. In practise we used the hyperfine fitting utility of
  the CLASS software package to simultaneously fit the five groups.
It is often difficult to measure the optical depth 
directly from the \nhthree\,(2,2) line profile as the 
signal-to-noise on the satellites is usually poor. Instead, $\tau_{\rm 2,2}$
may be calculated from the brightness temperatures of the (2,2) and
(1,1) main groups:
\begin{equation}
  {\rm \tau_{\,2,2} =
    -\frac{1}{f_{2,2}}\,ln\left[1-\frac{T_{B_{2,2}\,main}}{T_{B_{1,1}\,main}}\left(1-{\rm
	e}^{-\tau_{1,1}^{tot}f_{1,1}}\right)\right],}
\end{equation}
where f$_{1,1}$\,=\,0.502 and f$_{2,2}$\,=\,0.796 are the fractional
intensities in the (1,1) and (2,2) main groups, respectively, compared
to their satellites.
The rotational temperature is found from the optical depth
ratio $\tau_{2,2}/\tau_{1,1}$ via:
\begin{equation}
  {\rm T_{rot}=-41\,K/ln\,\left(\frac{9}{20}\,\frac{\tau_{2,2}}{\tau_{1,1}}\,\frac{\Delta\,\nu_{2,2}}{\Delta\,\nu_{1,1}}\right),}
\end{equation}
where $\tau_{\rm J,K}$ and $\Delta\nu_{\rm J,K}$ are the total optical
depth and the full-width half-maximum (FWHM) linewidth of the \nhthree\,(J,K) transition.
Finally, the rotational temperature may be converted directly to a
kinetic temperature T$_{\rm kin}$ via:
\begin{equation}\label{eqn:danby_trot}
  {\rm T_{rot} = \frac{T_{kin}}{[1+(T_{kin}/41\,K)\,ln(1+C_{23}/C_{21})]}}.
\end{equation}
\citet{Danby1988} have used large velocity gradient (LVG) models to
calculate the collisional coefficients C$_{23}$ and C$_{21}$,
for kinetic temperatures ranging from 5\,--\,300\,K. To derive
Equation~\ref{eqn:danby_trot} we assume that only the J,K\,=\,(1,1),
(2,1) and (2,2) transitions are involved. At low temperatures this
assumption is valid, however, above T$_{\rm rot}\,\sim$\,30\,K the
excitation of \nhthree~to higher energy levels becomes
non-negligible. This means that the above T$_{\rm rot}\,\sim$\,30\,K the
kinetic temperature found via comparison of the \nhthree\,(1,1) and
(2,2) lines becomes unreliable. \citet{Tafalla2004} have investigated
the accuracy of this method and find that kinetic temperatures derived
are accurate to better than 5 percent below 20\,K.

 
\subsection{Column density and core mass}\label{app:column_nh3}
If the optical depth and excitation temperature in any
(J,K) transition are known, the total column density of
emitting molecules may be calculated assuming local thermal
equilibrium (LTE) via the equation 
\begin{equation}\label{eqn:combined_column_total}
   {\rm N_{J,K} = \frac{8k\pi\nu_{J,K}^2}{A_{ul}hc^3}\int_{-\infty}^{\infty} T_b\,dv\,\left(\frac{\tau}{1-e^{-\tau}}\right)\,\frac{e^{(E_{J,K})/kT}}{g_{J,K}\,Q(T_{rot})}},
\end{equation}
where Q(T$_{\rm rot}$) is the partition function. The exact partition
function is the sum over all the levels
however, this can be approximated
by an exponential fit to the discrete values in the \citet{Pickett1998}
spectral line catalogue. For \nhthree~this is:
\begin{equation}
  {\rm Q(T) = 0.1266\,T^{1.48}.}
\end{equation}

\subsection{CO and \hcop}\label{app:co_col}
Physical conditions were calculated on a pixel-by-pixel basis in the
Mopra data. Using the {\it regrid} task in {\scriptsize MIRIAD}, the
CO data-cubes were first resampled to the same pixel-scale in RA, Dec
and velocity. The {\it maths} task was used to perform all subsequent
calculations. For each spatial pixel we calculated the optical depth
as a function of frequency using a modified version of the isotope
ratio 
\citep{Bourke1997}: 
\begin{equation}
  {\rm \frac{T_{MB}\,(\ceighteeno)}{T_{MB}\,(\thirteenco)}=
  \frac{1-e^{-(\tau_{_{\,18}}/X)}}{1-e^{-\tau_{_{\,13}}}}},
\end{equation}
In the above equation, T$_{\rm MB}$ is the peak main beam brightness
temperature and X is the abundance ratio of
[\thirteenco]\,/\,[\ceighteeno] assumed to be constant. During the
calculation pixels with insufficient emission were blanked, resulting
in a map of \thirteenco~optical depth with real values wherever
\ceighteeno~emission was detected. The expected
[\thirteenco/\ceighteeno] ratio in giant molecular clouds is
$\sim$\,11.5 (see \citealt{Goldsmith1997} and references therein),
however, we found that this value resulted in negative optical depths in some
parts of the map. The {\it lowest} abundance 
ratio that did not result in negative optical depths was
$\sim$\,16. This value was used in our calculations and should be
considered a lower limit. Optical depths ranged from
$0.1\leq\tau\leq4.4$. Using the \thirteenco~optical depth and peak
brightness temperature maps as input, we then constructed an
excitation temperature map via:
\begin{equation}
  {\rm T_{ex} = \frac{h\nu_u}{k}\,\left[ln\,\left(1+\frac{(h\nu_u/k)}{T_{13}/C_{\tau}+J_{\nu}(T_{bg})}\right)\right]^{-1}},
\end{equation}
where C${\rm _{\tau}=(1-e^{-\tau_{13}})}$. At this stage we blanked
pixels with spuriously excitation temperatures (T$_{\rm ex}\,<$4\,K
and T$_{\rm ex}\,>$40\,K) due to noisy \ceighteeno~data
which had passed our initial signal-to-noise cutoff. Assuming a
thermalised population with ${\rm T_{kin}=T_{ex}}$, a column density
map was be produced via Equation~\ref{eqn:combined_column_total}.
The partition function Q(T) for \thirteenco~has been measured exactly
at discrete values by \citet{Pickett1998} spectral line
catalogue. Intermediate values were determined via an exponential
fit, which for CO is:
\begin{equation}
  {\rm Q(T_{kin}) = 0.3968\,T_{kin}}.
\end{equation}
The CO column density map is presented in
Figure~\ref{fig:ngc3576_mopra_columns}-a. 

We calculated the column-density of \hcop~using the same procedure,
and a partition function of the form:
\begin{equation}
  {\rm Q(T_{kin}) = 0.4798\,T_{kin}}.
\end{equation}
The [\hcop]\,/\,[\hthirteencop] was set to 40 for these calculations.

\subsection{\ntwohp~and CS}\label{app:cs_col}
A slightly different approach was taken for the \ntwohp~data.
In a similar manner to \nhthree, the hyperfine structure in the spectrum of
\ntwohp\,(1\,--\,0) allows the direct determination of optical
depth. We fit all spectra in the
\ntwohp~data-cube using the HFS method in {\scriptsize CLASS} and
found that the \ntwohp~emission was 
consistent with it being optically thin ($\tau\ll\,0.1$) over the entire
cloud. Hence, no independent estimate could be made of the excitation
temperature. \ntwohp~and \nhthree~have similar critical densities and,
being nitrogen bearing molecules, are often detected under similar
conditions. We calculated the column
density of \ntwohp~assuming excitation temperatures equal to the
\nhthree~rotational temperatures. In practise, we smoothed the
T$_{\rm rot}$\,(\nhthree) map to the same resolution as the Mopra
\ntwohp~data. We then produced a column density map by combining this
T$_{\rm ex}$ map with the \ntwohp~integrated map via
Equation~\ref{eqn:combined_column_total} in the limit that $\tau \ll 0$. The
partition function for \ntwohp~was again interpolated from the values
in the \citet{Pickett1998} catalogue:
\begin{equation}
  {\rm Q(T_{kin}) = 4.198\,T_{kin}}.
\end{equation}

A column density map of CS was prepared in a similar manner, although
we note that our assumption of optically thin emission may not be
valid. Judging by the non-Gaussian shapes of some line profiles,
regions with strong emission may have significant optical depths.

\input tables/11358tbA1.tex

\end{document}

%% file: tables/11358tb1.tex
\begin{table*}
  \centering
  \begin{minipage}{130mm}
    \caption[Details of the Australia Telescope Compact Array observing runs.]{Details of the Australia Telescope Compact Array observing runs.}\label{tab:atcaobs}
    \begin{footnotesize}
    \begin{tabular}{lccccccc}
      \hline
	UT   & Array          & Beam\,$^{\alpha}$ & Transitions & Frequency & Bandwidth & No.       & K/(Jy/bm)\\
	Date & Config.  & Size\,(\arcsec)       &             & (GHz)          & (MHz) & Channels  &     \\
	\hline
	2003 Aug 28 & EW367 & 7.6\,$\times$\,6.2    & \nhthree\,(1,1)       & 23.694480 & ~~~8 &  512 & ~~46.02$^{\beta}$ \\
                    &       &                       & Continuum             & 23.694500 &  128 & ~~32 & -- \\
                    &       &                       & \htwoo\,(6--5)        & 22.235120 & ~~16 & ~~16 & -- \\
	2004 Jul 02 & 750D  & 6.6\,$\times$\,2.4    & \nhthree\,(4,4)       & 24.139417 & ~~~8 &  512 & 134.47 \\
                    &       &                       & Continuum             & 22.000000 &  128 & ~~32 & -- \\
	2005 Jul 17 & H75   & 26.8\,$\times$\,20.9  & \nhthree\,(1,1)       & 23.694480 &  ~~8 &  512 & ~~~~3.89 \\
                    &       &                       & \nhthree\,(2,2)       & 23.722630 &  ~~8 &  128 & ~~~~3.88\\
        \hline
    \end{tabular}      
    \begin{flushleft}
      $^{\alpha}$~Beam size excluding the 3\,km+ baselines to Antenna 6.\\
      $^{\beta}$~Jy to K conversion factor for the combined EW367 and
      H75 data is 17.58\, K/(Jy/bm).\\ 
    \end{flushleft}   
  \end{footnotesize}
  \end{minipage}
\end{table*}

%% file: tables/11358tb2.tex
\begin{table}
  \begin{center}
    \caption[Centre coordinates of the ATCA fields.]{~Centre
      coordinates of the ATCA fields corresponding approximately to
      the peak 1.2-mm emission.}\label{tab:ngc3576_coordinates}
      \begin{tabular}{lcc}
	\hline
	Field & RA      & DEC     \\
	      & (J2000) & (J2000) \\
	\hline
	S1 & 11$^h$12$^m$17.00$^s$ & -61$^d$17$^m$40.0$^s$\\
	S2 & 11$^h$12$^m$04.90$^s$ & -61$^d$18$^m$20.0$^s$\\
	S3 & 11$^h$11$^m$51.50$^s$ & -61$^d$18$^m$52.0$^s$\\
	S4 & 11$^h$11$^m$38.33$^s$ & -61$^d$19$^m$53.9$^s$\\
	S5 & 11$^h$11$^m$33.86$^s$ & -61$^d$21$^m$20.9$^s$\\
        \hline
      \end{tabular}      
  \end{center}
\end{table}

%% file: tables/11358tb3.tex
\begin{table*}
  \centering
      \caption[22\,GHz H$_2$O masers in NGC\,3576.]{~22\,GHz H$_2$O masers in NGC\,3576.}\label{tab:h2o_maser_coords}
      \begin{small}
      \begin{tabular}{llccccc}
	\hline
	Maser & RA      & Dec     & Velocity Range & Peak V$_{\rm LSR}$ & Peak & Reference.\\	
              &(J2000) & (J2000) & (\kms)         & (\kms)             & (Jy/beam) \\
	\hline  
	M1 & 11$^h$12$^m$16.72$^s$ & $-$61$^d$17$^m$47.0$^s$ & $-$27 to $-$19 & $-$24.3 & 5.1 & New \\
	M2 & 11$^h$12$^m$13.95$^s$ & $-$61$^d$17$^m$30.0$^s$ & $-$22 to $-$18 & $-$20.7 & 3.3 & New \\
	M3 & 11$^h$12$^m$00.76$^s$ & $-$61$^d$18$^m$31.0$^s$ & $-$41 to $+$15 & $-$25.5 & 6.5 & New \\
	M4 & 11$^h$11$^m$53.40$^s$ & $-$61$^d$18$^m$24.0$^s$ & $-$51 to $+$17 & $-$31.4 & 46.5 & Caswell 2004\\
	M5 & 11$^h$11$^m$50.07$^s$ & $-$61$^d$18$^m$51.0$^s$ & $-$110 to $+$12 & $-$101.5 & 69.3 & Caswell 2004 \\
	M6 & 11$^h$11$^m$39.80$^s$ & $-$61$^d$19$^m$41.0$^s$ & $-$27 to $-$20 & $-$20.9 & 7.8 & New \\
	M7 & 11$^h$11$^m$39.52$^s$ & $-$61$^d$20$^m$17.0$^s$ & $-$25 to $-$17 & $-$18.4 & 1.3 & New \\
	M8 & 11$^h$11$^m$34.52$^s$ & $-$61$^d$21$^m$16.0$^s$ & $-$31 to $-$3 & $-$27.7 & 24.8 & New \\
	M9 & 11$^h$11$^m$56.64$^s$ & $-$61$^d$19$^m$01.0$^s$ & $<-$134 to $-$121 & $-$129.9 & 670.5 & Caswell 2004 \\
	\hline
      \end{tabular}
      \end{small}
\end{table*}

%% file: tables/11358tb4.tex
\begin{table}
  \begin{center}
    \caption{~ATCA \nhthree~percentage flux detected.}\label{tab:missing_flux}
      \begin{tabular}{lc}
	\hline
	Field & Percentage Flux\\
	      & Detected \\
	\hline
	S1 & 94 \\
	S2 & 78 \\
	S3 & 80 \\
	S4 & 91 \\
	S5 & 70 \\
        \hline
      \end{tabular}      
  \end{center}
\end{table}

%% file: tables/11358tb5.tex
\begin{table*}
  \caption{~Properties of the \nhthree\,(1,1) cores found using the
  {\scriptsize FELLWALKER} routine in the ATCA data.}\label{tab:nh311_clumps}
  \begin{tabular}{lllc@{\hspace{3mm}}c@{\hspace{3mm}}c@{\hspace{3mm}}c@{\hspace{3mm}}c@{\hspace{2mm}}c@{\hspace{2mm}}c@{\hspace{3mm}}c@{\hspace{3mm}}c@{\hspace{3mm}}c@{\hspace{3mm}}c@{\hspace{3mm}}c}
    \hline
    ID & RA  & Dec & $\Delta$RA$^{\alpha}$ &  $\Delta$Dec$^{\alpha}$ & $\Delta$V & T$_{\rm MB}$ & $\sum$T$_{\rm MB}$\,dv & N$_{\nhthree}$ & M$_{\rm LTE_1}$$^{\beta}$ & M$_{\rm LTE_2}$$^{\gamma}$ & M$_{\rm vir_1}$$^{\delta}$ & M$_{\rm vir_2}$$^{\dagger}$ & T$_{\rm kin}$ & $\tau$\\
       & (J2000) & (J2000) & (\arcsec) & (\arcsec) & (km/s) & K & (K\,km/s) & ($\times\,10^{15}$cm$^{-2}$) & (\msun) & (\msun) & (\msun) & (\msun) & K & \\
    \hline
    1 & 11:12:21.30 & -61:17:11.5 & 12.4 & 4.9 & 1.3 & 20.4 & 2803.6 & 20.8 & 82 & 79 & 18 & 11 & 15 & 1.5 \\
    2 & 11:12:16.70 & -61:17:39.0 & 10.0 & 8.9 & 1.9 & 23.0 & 4989.6 & 36.7 & 144 & 166 & 41 & 25 & 20 & 0.7 \\
    3 & 11:12:14.18 & -61:17:21.2 & 6.4 & 8.4 & 2.4 & 18.5 & 2427.1 & 18.5 & 73 & 168 & 53 & 32 & 22 & 0.7 \\
    4 & 11:12:08.91 & -61:18:02.9 & 6.8 & 6.1 & 1.2 & 6.1 & 485.6 & 3.9 & 15 & 15 & 11 & 7 & 33 & 0.1\\
    5 & 11:12:05.25 & -61:18:49.3 & 18.0 & 8.8 & 1.7 & 5.6 & 1538.2 & 11.3 & 44 & 44 & 48 & 29 & 19 & 0.1 \\
    6 & 11:12:04.76 & -61:18:24.8 & 7.1 & 1.6 & 0.8 & 8.5 & 131.7 & 1.2 & 5 & 5 & 3 & 2 & 47 & 0.1 \\
    7 & 11:12:01.53 & -61:18:58.7 & 5.8 & 4.0 & 1.0 & 6.0 & 232.1 & 1.8 & 7 & 7 & 6 & 3 & 25 & 0.1 \\
    8 & 11:11:52.45 & -61:18:30.2 & 4.2 & 3.7 & 0.6 & 23.9 & 185.2 & 1.6 & 6 & 47 & 2 & 1 & 31 & 0.1 \\
    9 & 11:11:51.91 & -61:18:22.2 & 7.3 & 4.8 & 1.0 & 10.0 & 668.9 & 5.8 & 23 & 171 & 8 & 5 & 31 & 0.1\\
    10 & 11:11:48.39 & -61:18:54.1 & 2.8 & 3.7 & 1.2 & 28.4 & 809.4 & 7.0 & 27 & 91 & 6 & 3 & 40 & 0.1 \\
    11 & 11:11:46.05 & -61:18:43.4 & 8.9 & 9.4 & 1.7 & 17.8 & 3990.0 & 32.5 & 128 & 426 & 32 & 19 & 33 & 0.1\\
    12 & 11:11:45.66 & -61:18:52.0 & 9.5 & 9.3 & 0.9 & 9.4 & 1051.4 & 8.6 & 34 & 112 & 10 & 6 & 35 & 0.1 \\
    13 & 11:11:41.86 & -61:19:32.4 & 12.8 & 7.7 & 1.9 & 23.2 & 4935.5 & 37.8 & 148 & 62 & 44 & 27 & 26 & 0.4 \\
    14 & 11:11:39.38 & -61:19:44.2 & 8.0 & 5.6 & 1.5 & 30.2 & 4757.6 & 35.5 & 139 & 59 & 20 & 12 & 23 & 0.7 \\
    15 & 11:11:38.90 & -61:20:15.6 & 3.3 & 4.1 & 0.8 & 23.3 & 647.6 & 4.8 & 19 & 16 & 3 & 2 & 21 & 1.2 \\
    16 & 11:11:38.11 & -61:21:55.7 & 3.9 & 4.6 & 1.6 & 4.3 & 222.1 & 1.8 & 7 & 14 & 13 & 8 & -- & 0.1 \\
    17 & 11:11:37.94 & -61:21:06.5 & 1.3 & 2.8 & 0.8 & 85.9 & 595.2 & 4.5 & 18 & 35 & 1 & 1 & 15 & 1.2 \\
    18 & 11:11:37.92 & -61:19:35.1 & 6.6 & 4.4 & 1.0 & 17.5 & 975.2 & 7.3 & 29 & 12 & 7 & 4 & 24 & 1.0 \\
    19 & 11:11:36.49 & -61:20:07.0 & 7.6 & 12.3 & 1.7 & 25.9 & 6502.5 & 47.9 & 188 & 166 & 36 & 22 & 17 & 1.0 \\
    20 & 11:11:35.62 & -61:21:30.0 & 7.9 & 2.6 & 0.6 & 20.5 & 422.7 & 3.2 & 12 & 25 & 3 & 2 & 15 & 2.2 \\
    21 & 11:11:33.11 & -61:21:11.6 & 13.2 & 8.4 & 2.4 & 27.9 & 8117.8 & 65.7 & 258 & 516 & 78 & 47 & 15 & 1.9 \\
    22 & 11:11:32.84 & -61:19:53.0 & 1.2 & 0.3 & 0.8 & 321.3 & 231.1 & 1.7 & 7 & 6 & 1 & 0 & 17 & 1.0 \\
    23 & 11:11:27.13 & -61:21:07.3 & 5.6 & 2.0 & 0.7 & 32.3 & 517.0 & 3.9 & 15 & 30 & 2 & 1 & 16 & 1.6 \\
    24 & 11:11:24.25 & -61:21:18.7 & 5.2 & 6.2 & 0.8 & 4.4 & 109.9 & 0.9 & 3 & 7 & 5 & 3 & -- & 0.1 \\
    25 & 11:11:24.03 & -61:21:25.3 & 6.5 & 6.6 & 0.9 & 8.6 & 534.0 & 4.2 & 17 & 33 & 7 & 4 & -- & 0.1 \\
    \hline
  \end{tabular}
  \begin{flushleft}
    $^{\alpha}$ Deconvolved full-width half-maximum (FWHM) core size
    asumming an average beam FWHM of 11.2\arcsec.

    $^{\beta}$ LTE-mass derived from the integrated
    intensity of \nhthree\,(1,1) assuming a relative abundance
    [\nhthree]\,/\,[H$_2$]\,=\,$3\,\times\,10^{-8}$. 
      
    $^{\gamma}$ Corrected LTE-mass, derived using the
    [\nhthree]\,/\,[H$_2$] abundance ratios calculated in
    Section~\ref{sec:clump_masses} via a comparison between
    \nhthree~and 450\,\micron~p-ArT\'eMiS data.
      
    $^{\delta}$ M$_{\rm vir}$ derived assuming a constant density profile.

    $^{\dagger}$ M$_{\rm vir}$ derived assuming a 1/r$^2$ density profile.
  \end{flushleft}
\end{table*}

%% file: tables/11358tb6.tex
\begin{table*}
  \centering
  \begin{minipage}{150mm}
  \caption{~Comparison of masses derived from p-ArT\'eMiS 450\micron~and ATCA \nhthree\,(1,1) emission.}\label{tab:clump_mass}
  \begin{tabular}{lcccccccc}
    \hline
    Clump & RA         & Dec         &  Mass          & Mass          & [\nhthree]\,/\,[H$_2$]$^{\alpha}$ & Average     & Aliases           \\
    ID    & (J2000)    & (J2000)     &  450\micron    & \nhthree~LTE  & Abundance                        & T$_{\rm dust}$ & \citep{Andre2008} \\
          &            &             &  (\msun)       & (\msun)       & \\
    \hline
    S1-a  & 11:12:21.09 & -61:17:09.50 & ~~89           & ~~91          &  $3.1\times10^{-8}$              & 15             & -- \\
    S1-b  & 11:12:17.21 & -61:17:43.50 &  161           &  140          &  $2.6\times10^{-8}$              & 21             & S1-M1 \\  
    S1-c  & 11:12:14.43 & -61:17:17.50 &  193           & ~~82          &  $1.3\times10^{-8}$              & 23             & S1-M2 \\
    S3-a  & 11:11:53.05 & -61:18:31.37 &  357           & ~~51          &  $0.4\times10^{-8}$              & 40             & S3-M4 \\
    S3-b  & 11:11:45.54 & -61:18:45.27 &  682           &  211          &  $0.9\times10^{-8}$              & 38             & S3-C3 \\
    S4-a  & 11:11:39.69 & -61:19:43.18 &  161           &  383          &  $7.1\times10^{-8}$              & 24             & S4-M6 \\
    S4-b  & 11:11:36.91 & -61:20:03.13 &  247           &  279          &  $3.4\times10^{-8}$              & 19             & -- \\
    S5    & 11:11:34.66 & -61:21:11.09 &  712           &  339          &  $1.4\times10^{-8}$              & 15             & S5-M8 \\

    \hline
  \end{tabular}
  \begin{flushleft}
    \begin{footnotesize}
      $^{\alpha}$ Relative abundance of \nhthree~to H$_2$ via 
      a direct comparison of clump masses derived from \nhthree~and
      450\,\micron~emission.\\ 
    \end{footnotesize}
  \end{flushleft}
  \end{minipage}
\end{table*}

%% file: tables/11358tbA1.tex
\begin{table*}
  \centering
  \begin{minipage}{110mm}
    \caption{~Molecular constants\,$^{\alpha}$.}\label{tab:transitions3}
    \begin{tabular}{lccccc}
      \hline
      Species & Transition    & Frequency & E$_{\rm u}$/k & A$_{\rm ul}$ & n$_{\rm crit}\,^{\beta}$ \\
              & (J$_{\rm K}$) & (GHz)     & (K)           & (s$^{-1}$)   & (\cmmthree) \\
      \hline
      \hcop \rule{0pt}{4mm}         & $1\rarr0$          & ~~89.188526 & ~~4.280  & ~~3.020$\times 10^{-5}$  & 3.0$\times 10^{5}$\,$^{\dagger}$\\
      \hthirteencop \rule{0pt}{4mm} & $1\rarr0$          & ~~86.754330 & ~~4.160  & ~~2.800$\times 10^{-5}$  & 3.0$\times 10^{5}$\,$^{\dagger}$\\
      \thirteenco \rule{0pt}{4mm}   & $1\rarr 0$         &  110.210353 & ~~5.288 & 6.389$\times 10^{-8}$ & $6.5\times\,10^2$\,$^{\dagger}$ \\
      \ceighteeno                   & $1\rarr 0$         &  109.782173 & ~~5.267 & 4.961$\times 10^{-7}$ & --              \\
      CS                            & $2\rarr 1$         & ~~97.980950 & ~~7.052 & 1.703$\times 10^{-5}$ & $2.0\times\,10^5$ \\
      \ntwohp \rule{0pt}{4mm}       & $1_{1}\rarr 0_{1}$  & ~~93.171880 & ~~4.471 & 3.654$\times 10^{-5}$ & 2$\times 10^{5}$\,$^{un}$\\
                                    & $1_{2}\rarr 0_{1}$  & ~~93.173700 & --      & --                    & -- \\
                                    & $1_{0}\rarr 0_{1}$  & ~~93.176130 & --      & --                    & -- \\
      \nhthree \rule{0pt}{4mm}      & $1_{1}\rarr 1^{\prime}_{1^{\prime}}$ & ~~23.694480 & 17.215  &  1.67$\times 10^{-7}$ & 2$\times 10^{4}$\,$^{sw}$\\
                                    & $2_{2}\rarr 2^{\prime}_{\,2^{\prime}}$ & ~~23.722630 & 63.866  &  2.29$\times 10^{-7}$ & \\
      \hline
    \end{tabular}
    \begin{flushleft}
      \begin{footnotesize}
        $^{\alpha}$~~~Unless otherwise noted, all
        values quoted here are sourced from the NASA Jet Propulsion Lab
        (JPL) spectral line database \citep{Pickett1998}.\\
    
        $^{\beta}$~~~Critical density values marked with a
        $\dagger$ are calculated from the formula ${\rm n_{crit} =
          A_{ul}/\gamma_{_{H_2}}}$, where ${\rm \gamma_{_{H_2}}}$ is the
        `standard' collisional rate coefficient for H$_2$, assumed equal
        to 10$^{10}$ ${\rm cm^3.s^{-1}.H_2}$-${\rm molecules^{-1}}$. Other
        values are referenced as follows: $^{un}$~\citet{Ungerechts1997}, 
	$^{sw}$\,\citealt{Swade1989}.
      \end{footnotesize}
    \end{flushleft}
  \end{minipage}
\end{table*}